\documentclass[runningheads]{llncs}
\usepackage[T1]{fontenc}
\usepackage{graphicx}

\usepackage{hyperref}
\usepackage{booktabs} 
\usepackage{siunitx} 
\usepackage{caption} 
\usepackage{geometry} 
\usepackage{cite}
\usepackage{amsmath}
\usepackage{amsfonts}
\usepackage{adjustbox}
\usepackage{algorithmic}
\usepackage{tabularray}
\usepackage{balance}

\usepackage{textcomp}
\usepackage{longtable}
\usepackage{tcolorbox}

\usepackage{xcolor}
\usepackage{subcaption} 

\usepackage{mdframed}

\newmdenv[
  backgroundcolor=gray!10,
  linecolor=black,
  linewidth=0.5pt,
  skipabove=10pt,
  skipbelow=10pt,
  innertopmargin=6pt,
  innerbottommargin=6pt,
  leftmargin=0pt,
  rightmargin=0pt,
]{infobox}

\begin{document}
\title{Can We Unmask the Underground? Detecting and Predicting Hidden Forum Interactions}
%
%
\author{Abdoul Nasser Hassane Amadou\and
Imane Fouad\and
Anas Motii}

\institute{College of Computing, Mohammed VI Polytechnic University (UM6P), Ben Guerir, Morocco\\
\texttt{\{abdoul.nasser, imane.fouad, anas.motii\}@um6p.ma}}
\maketitle              
\begin{abstract}
Cybercriminal underground forums enable anonymous collaboration, allowing users to trade illicit tools, discuss vulnerabilities, and distribute stolen data. Driven by shared interests and specialized skills, users on these platforms organize into distinct threat communities. However, identifying these communities is challenging due to their dynamic and opaque structures; traditional graph-based methods typically isolate dominant groups while overlooking smaller, hidden subgroups.
This paper introduces HADES, an unsupervised framework designed to detect both dominant and hidden communities in underground forums. The framework models users based on their textual interactions and leverages pretrained language models to generate semantic embeddings that encode latent behavioral and thematic patterns.  By clustering users based on semantic similarity and assigning topic labels to the resulting clusters, HADES identifies specific threat communities to support Cyber Threat Intelligence (CTI). The framework was evaluated on three major underground forums: HackForums, Cracked, and BreachForums. Results demonstrate that BERT embeddings consistently outperform alternative baselines, improving cluster coherence and achieving higher silhouette scores. Across these platforms, HADES identified dozens of distinct communities, effectively isolating small, specialized subgroups (fewer than 100 users) that evade traditional detection. Furthermore, because shared thematic interests frequently precede explicit structural connections, tracking these semantic patterns enables the framework to anticipate the formation of threat communities up to a year before they become detectable by traditional graph-based methods.
\end{abstract}
\section{Introduction}
Threat actors operating in underground forums pose significant risks to critical sectors, including finance, healthcare, and energy, with global cybercrime costs projected to reach \$23 trillion by 2027~\cite{sentinelone2026}. These forums serve as hubs for cybercriminal collaboration, where users self-organize into threat communities—dynamic subgroups connected by shared interests, complementary skills, or mutual malicious objectives~\cite{samtani2015exploring}. Detecting such communities enhances Cyber Threat Intelligence (CTI) by enabling focused analysis of high-risk groups engaged in activities such as malware distribution, data trading, or coordinated attacks.

Existing community detection approaches primarily rely on social network graphs derived from user interactions. Early methods identified disjoint communities~\cite{huang2019topic,pete2020social,pourhabibi2021darknetexplorer}, while more recent work detects overlapping groups~\cite{MANATOVA2024114271,Rios2012KDD}. However, these structural methods depend on dense, observable connections and consequently fail to detect \emph{hidden communities}: user groups characterized by sparse or deliberately concealed interactions. Such groups—including clandestine actors limiting public interactions for operational security and low-engagement clusters—exhibit weak connectivity and low modularity~\cite{li2024comprehensive}, causing them to evade conventional detection.

To address this gap, we analyze user-generated textual content to uncover shared semantic and behavioral patterns that persist even when structural connections are absent. However, the adversarial and unstructured nature of underground forum content poses significant challenges for reliable community detection. This work addresses three research questions: \textbf{(RQ1)}~How can we define and characterize hidden communities in underground forums? \textbf{(RQ2)}~How can we effectively detect these communities by analyzing user-generated semantic content? \textbf{(RQ3)}~Can semantic analysis identify emerging communities based on shared interests before traditional graph-based approaches detect structural connections?

We propose HADES, an unsupervised framework that detects both dominant and hidden communities without requiring labeled data. HADES constructs textual user profiles from forum interactions, generates semantic embeddings using pretrained language models, clusters users via HDBSCAN~\cite{mcinnes2017hdbscan}, and assigns descriptive labels to each cluster using a large language model. Because the framework operates on semantic similarity rather than structural connectivity, it generalizes across diverse underground forums and identifies cohesive groups even within sparsely connected networks.

We evaluate multiple embedding models to optimize clustering performance across diverse underground forums, using a multi-metric framework to comprehensively assess clustering quality. Experiments on real-world forum data identify several hidden communities that conventional graph-based methods fail to detect. These communities demonstrate meaningful, albeit sparse, structures that deepen our understanding of underground ecosystems.

To the best of our knowledge, this study is the first to investigate hidden community detection in underground forums using solely user textual interactions to identify both hidden and dominant groups. The primary contributions of this paper are:

\begin{itemize}
    \item We propose HADES, an unsupervised framework for detecting both hidden and dominant communities. The framework constructs user profiles from textual interactions, generates semantic embeddings, clusters users, and automatically labels the resulting clusters. Because it requires no labeled data, the approach generalizes across diverse underground forums.

    \item We evaluate the framework using large-scale, real-world data from underground forums. Through a multi-metric approach that combines intrinsic clustering quality measures with qualitative case studies, we demonstrate that HADES accurately identifies coherent user communities.
    
    \item We demonstrate the framework's early-detection capabilities through in-depth case studies. Notably, HADES identifies hidden groups discussing hacking and security on HackForums nearly a year before traditional graph-based methods detect structural connections, providing early insights that support proactive CTI. 
    
\end{itemize}
We release the source code at \url{https://github.com/osteoner/HADES} to facilitate reproducibility and enable the application of our methodology to other datasets.

\section{Related Works}

Understanding how threat actors organize within underground forums has attracted growing research attention, driven by the need to support proactive CTI. Early studies focused on finding influential or central actor groups within social networks. Pete et al.\cite{pete2020social} applied centrality-based social network analysis to identify key members, such as vendors and moderator groups. Huang et al.\cite{huang2021hackerrank} used topic modeling and graph analysis to filter for high-ranking users. Similarly, Huang et al.\cite{huang2016exploring} combined topic-based clustering with social network analysis, although their method ultimately relied on influence scores to identify key members. More recently, Amadou et al.~\cite{amadou2024hc,amadou2024eurekha} advanced key hacker identification by first integrating centrality measures with topic modeling and subsequently combining large language models with graph neural networks to jointly model user features and interaction structure. 

A primary limitation of these influence-centric methods is their bias toward highly active users, which often leads them to overlook smaller, low-profile, or covert groups.

Recent approaches apply community detection algorithms directly to forum network graphs, representing users as nodes and interactions as edges. These methods include using a surprise function on multigraphs to find unexpected community structures~\cite{pourhabibi2021darknetexplorer} and applying modularity-based algorithms to identify overlapping or influential groups~\cite{Rios2012KDD}. More advanced models, such as heterogeneous graph attention networks, have also been used to identify hacker groups~\cite{XU2022848,XU2024111587}. To systematically evaluate these approaches, Manatova et al.~\cite{MANATOVA2024114271} compared 18 algorithms across five prominent underground forums. However, a key limitation of these structural methods is their dependence on dense, observable network connections. Consequently, they often fail to detect communities whose interactions are deliberately sparse, fragmented, or designed to evade detection. 

To address these limitations, we propose a semantics-driven framework to detect both dominant and hidden communities. Rather than relying on structural connectivity, this method analyzes semantic content, specifically user discussions and profile attributes. The approach consists of four stages: (1) modeling users as textual sequences derived from their posts and profiles; (2) using a pretrained language model to generate semantic embeddings for each user; (3) clustering users based on embedding similarity; and (4) automatically generating descriptive labels for each identified community. By capturing semantic relationships among users, this content-centric approach identifies cohesive groups even within the sparsely connected networks characteristic of underground forums.

\section{System Design}
This section presents the HADES framework, which consists of three stages: (i) modeling users through their textual interactions to generate comprehensive semantic embeddings using pretrained language models; (ii) clustering users into structurally and semantically cohesive communities; and (iii) identifying high-threat actor groups to prioritize investigations for CTI analysts.

\subsection{Problem Formulation}
The sparsity and concealed nature of underground forums present distinct challenges for community detection, particularly when relying on traditional graph-based methods. Let $G = (V, E)$ denote an underground forum network, where $V$ is a set of $n$ nodes (e.g., users) and $E$ is a set of $m$ edges (e.g., user interactions). The corresponding adjacency matrix $A$ has entries $A_{ij} \in \{0,1\}$ indicating whether an edge exists between nodes $i$ and $j$. A community structure can be represented as $\mathcal{C} = \{c_1, c_2, \ldots, c_K\}$, where each $c_k = (V_k, E_k)$ is a comparatively dense subgraph of $G$. To quantify the quality of these communities, researchers often employ objective functions $F:(G, c_k) \mapsto \mathbb{R}$, such as \emph{modularity}~\cite{newman2004finding} or \emph{conductance}~\cite{868688}. Modularity, for instance, compares the observed fraction of intra-community edges to the expected fraction under a random null model that preserves the degree distribution. 
The modularity $Q$ of a network is calculated as: 
$Q = \frac{1}{2m} \sum_{ij} \left[ A_{ij} - \frac{k_i k_j}{2m} \right] \delta(c_i, c_j)$
where $m$ denotes the total number of edges in the network, $k_i$ and $k_j$ are the degrees of nodes $i$ and $j$, and $\delta(c_i, c_j) = 1$ if nodes $i$ and $j$ belong to the same community, and $0$ otherwise. The modularity score $Q$ lies in the range $[-1, 1]$, with larger values indicating a stronger community structure. Conductance, conversely, evaluates the separation of a community from the rest of the graph, typically yielding values between $0$ and $1$, where lower values indicate better separation.

\begin{figure}[htbp]
    \centering
    \includegraphics[width=0.8\linewidth]{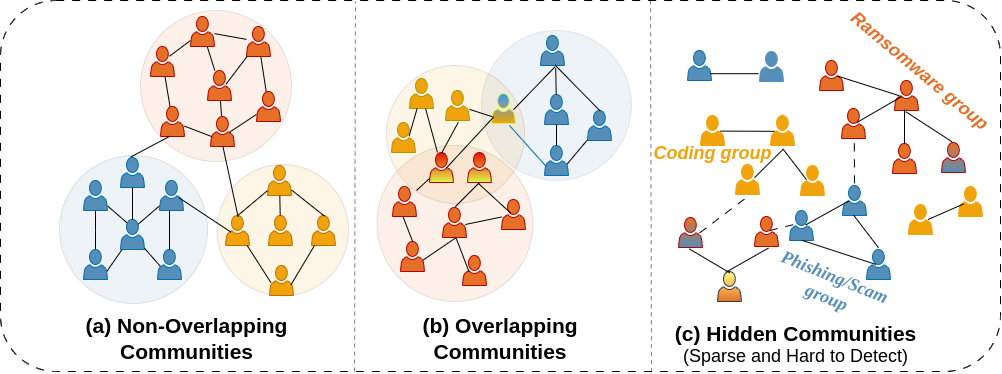}
    \caption{\textbf{Conceptual comparison of community structures in underground forum networks.} Traditional graph-based methods identify dominant communities characterized by dense, observable interactions (a, b). In contrast, hidden communities (c) exhibit sparse structural connections but remain united by shared semantic interests and illicit activities (e.g., ransomware, phishing), which often evade detection by structural methods.}
    \label{fig:mode-community}
\end{figure}

Although these measures capture structural cohesion, they often fail in highly sparse networks like underground forums, where user interactions are irregular, and many accounts exhibit minimal activity. As illustrated in Fig.~\ref{fig:mode-community}, dominant communities are large, densely connected groups of users who interact frequently and openly share resources. Their visible connections and high activity levels make them easily identifiable via standard graph-based analysis. In contrast, hidden communities are challenging to detect. Members of these groups may interact infrequently or maintain a low profile, yet they remain united by shared interests, behavioral patterns, or discussion topics. Their sparse connections allow them to blend into the broader forum or overlap with dominant groups, successfully evading traditional detection methods. For example, Fig.~\ref{fig:mode-community} shows a phishing community that appears as two disconnected groups under standard analysis due to limited direct interactions. In reality, these users form a single, cohesive community united by shared phishing activities.

The objective of HADES is to detect these hidden communities, denoted as $c_k$, by leveraging shared interests, semantic topics, and profile attributes rather than relying solely on structural edges. Formally, we approach this challenge by processing a continuous stream of $n$ underground forum users, $\mathcal{U} = \{u_1, u_2, \ldots, u_n\}$, to achieve four main objectives. First, we \textbf{model user profiles}, defining $\mathcal{S} = \{s_1, s_2, \ldots, s_n\}$, where each profile $s_i$ aggregates the sequence of actions, metadata, and content produced by user $u_i$. Second, we \textbf{embed users}, generating $\mathcal{E} = \{e_1, e_2, \ldots, e_n\}$, where each $d$-dimensional embedding $e_i \in \mathbb{R}^d$ is derived from $s_i$ to capture semantic and interactional features. Third, we \textbf{cluster users}, forming $\mathcal{C} = \{c_1, c_2, \ldots, c_K\}$, where each cluster $c_k$ groups the embeddings of users discussing specific threats, tools, or illicit services. Finally, we \textbf{label clusters}, computing $\mathcal{R} = \text{label}\big(\{(c_1, S_1), (c_2, S_2), \ldots, (c_K, S_K)\}\big)$, where $S_k$ represents the aggregated semantic context of cluster $c_k$. This produces a prioritized list of descriptive tags to guide CTI analysts.

This pipeline is expressed as:  
\[
\mathcal{S} = f_{\text{model}}(\mathcal{U}),\ 
\mathcal{E} = f_{\text{embed}}(\mathcal{S}),\ 
\mathcal{C} = f_{\text{cluster}}(\mathcal{E}),\ 
\mathcal{R} = f_{\text{label}}(\mathcal{C})
\]
where $f_{\text{model}}$ constructs user profiles, $f_{\text{embed}}$ generates vector representations, $f_{\text{cluster}}$ discovers user communities, and $f_{\text{label}}$ assigns interpretable semantic tags to clusters to prioritize threats for analyst review.

\subsection{System Overview}
\begin{figure}
    \centering
    \includegraphics[width=1\linewidth]{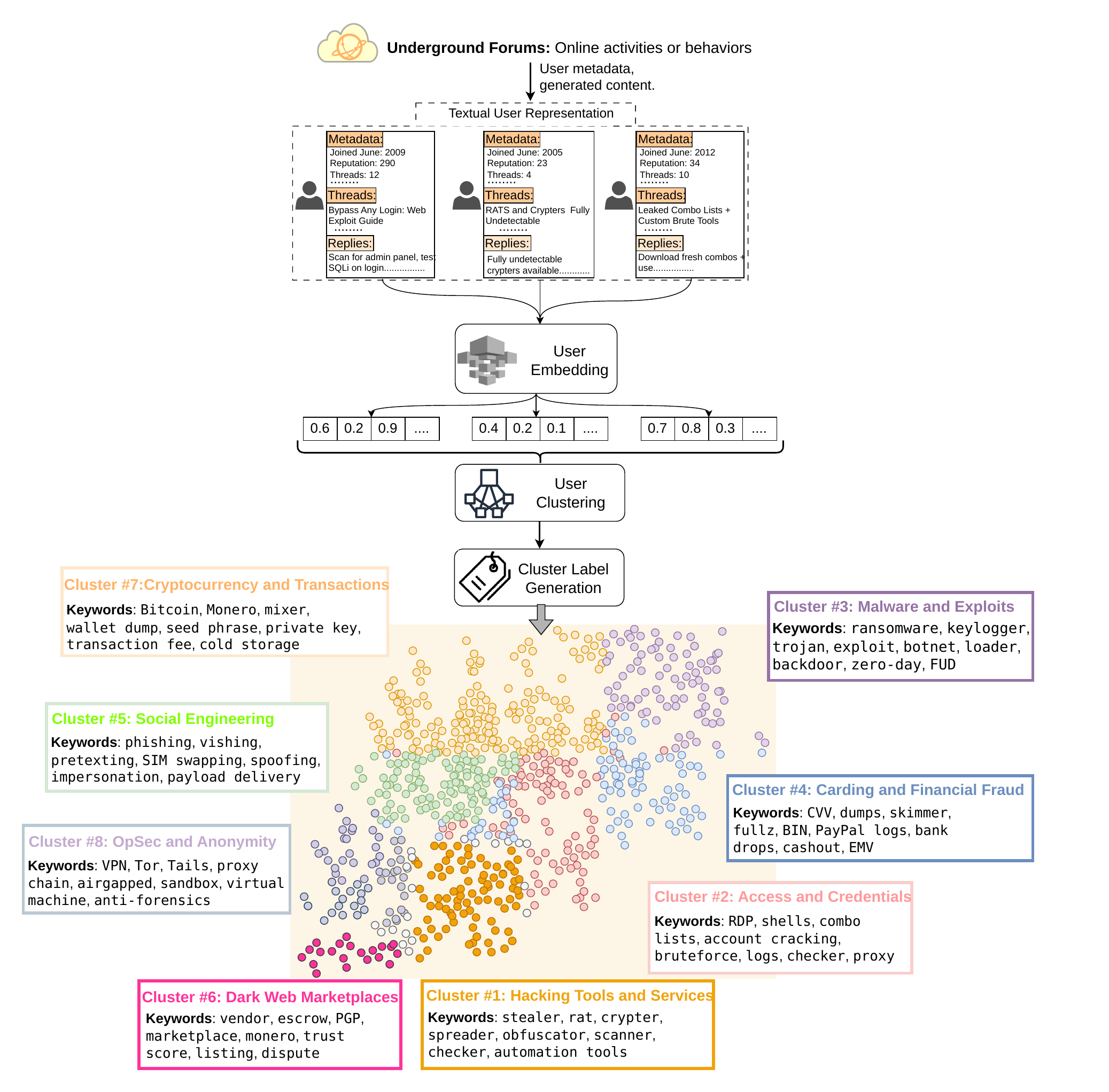}
    \caption{\textbf{Architecture of the HADES framework.} The pipeline extracts raw forum data to construct textual user representations. These sequences are embedded using pretrained language models and clustered via HDBSCAN to group users with similar behaviors. A large language model (Qwen2.5) evaluates representative posts from each cluster to generate descriptive thematic labels.}
    \label{fig:system-overview}
\end{figure}

This section details the architecture of HADES, illustrated in Figure~\ref{fig:system-overview}. The system pipeline comprises four main components: (1) a data selection and preprocessing module (Sec.~\ref{section:dataset}) that extracts data from prominent underground forums to capture a broad spectrum of cybercriminal activity (e.g., hacking, malware, and marketplaces) and removes noise such as HTML tags and URLs; (2) a user representation module (Sec.~\ref{section:user_representation}) that models forum members as comprehensive textual profiles by aggregating metadata and historical content to capture latent behavioral patterns; (3) an embedding and clustering module (Sec.~\ref{section:clustering}) that applies a pretrained language model and HDBSCAN~\cite{mcinnes2017hdbscan} to discover both dominant and hidden communities; and (4) a cluster labeling module (Sec.~\ref{section:label}) that prompts an LLM to extract representative content and categorize each identified community. Finally, Section~\ref{section:evaluation} details the intrinsic metrics used to evaluate system performance.

\subsection{Dataset and Preprocessing}
\label{section:dataset}

\noindent \textbf{Data Source}
We used the CrimeBB dataset~\cite{Pastrana2018CrimeBB}, an established archive of cybercriminal forum activity~\cite{paracha2023sus, cabrero2021methodology, pete2022postcog, wilson2024identifying}. The corpus contains approximately 110 million posts authored by 6 million members across 36 forums, providing a comprehensive empirical basis for analyzing global cybercriminal communities.

\noindent \textbf{Underground Forum Selection}
From this corpus, we selected three English-language forums based on their diverse user behaviors, varying levels of technical sophistication, and roles as hubs for sharing exploits, source code, and illicit tools: \textit{HackForums}\footnote{\texttt{https://hackforums.net/}} focuses on malware, social engineering, and fraud techniques; \textit{Cracked}\footnote{\texttt{https://cracked.net/}} operates as a marketplace for cracked software and stolen credentials, emphasizing account cracking and fraud; and \textit{BreachForums}\footnote{\texttt{https://breachforums.st/}}, the successor to RaidForums, serves as a primary platform for distributing leaked databases and exploits. Table~\ref{tab:forum_datasets} summarizes the key characteristics of these forums. By hosting interactions that range from technical knowledge sharing to explicit transactions for illicit services, these platforms provide robust environments for studying group structures within cybercriminal networks.

\noindent \textbf{Data Preprocessing}
Raw forum data is inherently noisy and unstructured. We applied a standardized preprocessing pipeline~\cite{maharana2022review} to reduce noise and normalize the text before embedding and clustering. Specifically, we (1) removed HTML markup and other non-informative text; (2) replaced URLs, cited posts, quotations, and source code with canonical tokens (\texttt{URL}, \texttt{CITE}, \texttt{QUOTE}, \texttt{CODE}); and (3) filtered out non-Latin characters. This pipeline reduces the dimensionality of the feature space, improving both computational efficiency and clustering interpretability~\cite{petukhova2022textcl}.

\begin{table}[h]
\centering
\caption{\textbf{Characteristics of the selected underground forum datasets.} The corpus spans three prominent platforms (HackForums, Cracked, and BreachForums) with varying levels of technical sophistication, covering millions of posts and threads to ensure robust cross-platform evaluation.}
\label{tab:forum_datasets}
\begin{tabular}{lrrrl}
\toprule
\textbf{Forum} & \textbf{Posts} & \textbf{Threads} & \textbf{Members} & \textbf{Time Span} \\
\midrule
HackForums   & 42,474,325 & 4,148,196 & 716,058  & Jan 2007 -- Sep 2024 \\
Cracked      & 2,977,800  &   419,517 & 897,760  & Apr 2018 -- Jun 2023 \\
BreachForums &   737,922  &    34,412 & 119,260  & Mar 2022 -- Sep 2024 \\
\bottomrule
\end{tabular}%
\end{table}

\subsection{Textual User Representation}
\label{section:user_representation}
\begin{figure*}
    \centering
    \includegraphics[width=0.98\linewidth]{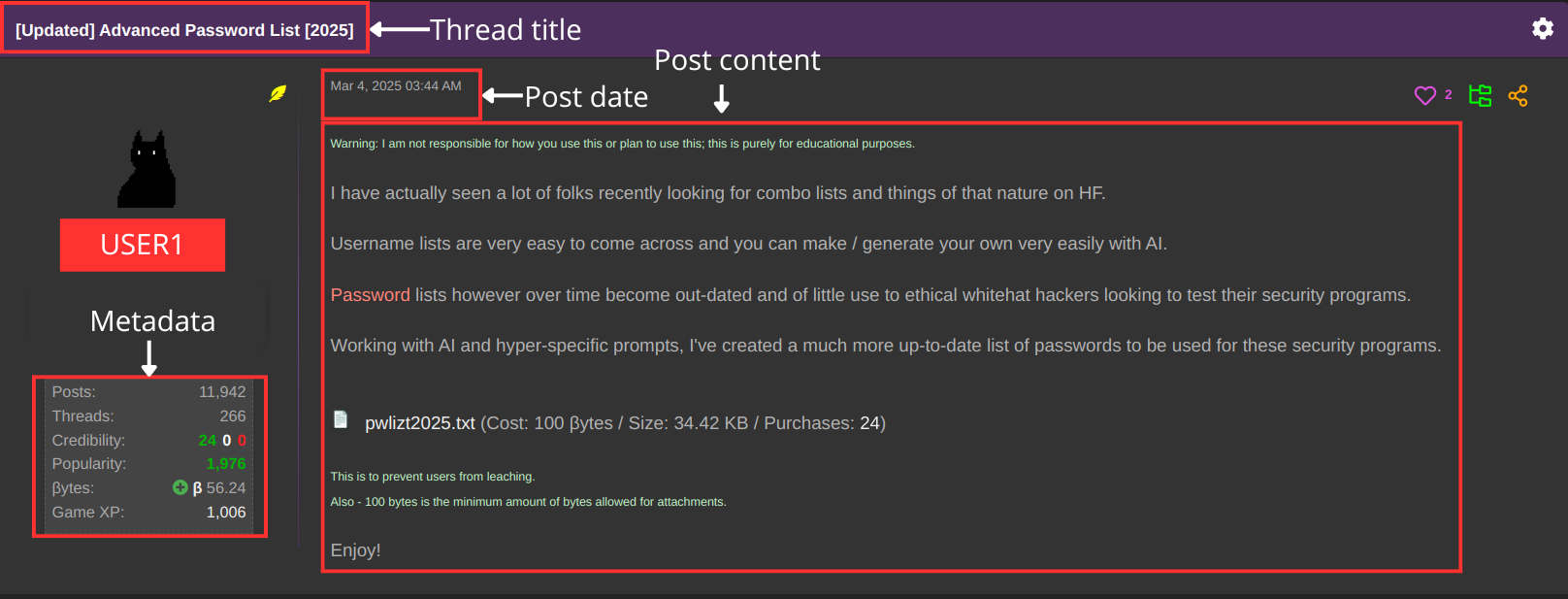}
    \caption{\textbf{Anonymized example of user interactions on HackForums.} The snapshot illustrates the unstructured nature of forum data.}
    \label{fig:user_activity}
\end{figure*}

To uncover latent behavioral patterns and characterize members within underground forums, we represent each individual as a cohesive textual sequence. As illustrated in Fig.~\ref{fig:user_activity}, user profiles contain metadata and records of various activities, such as initiating threads, replying to posts, and sending direct messages. We define each user through two primary components: \textbf{Metadata} and \textbf{Content}. Metadata includes \emph{usernames} (identifiers that often indicate specialization or experience, e.g., \textit{CardMaster2024}, \textit{MalwareExpert}); \emph{reputation scores} (forum-specific trust indicators derived from peer feedback, resource-quality ratings, moderator endorsements, and transaction histories); and \emph{activity metrics} (such as the number of contributed posts and initiated threads). Content comprises \emph{threads} (complete discussions, including tutorials, tool sharing, marketplace advertisements, and operational planning) and \emph{posts} (message-level contributions reflecting technical expertise and communication patterns).

\begin{figure}
    \centering
    \includegraphics[width=1\linewidth]{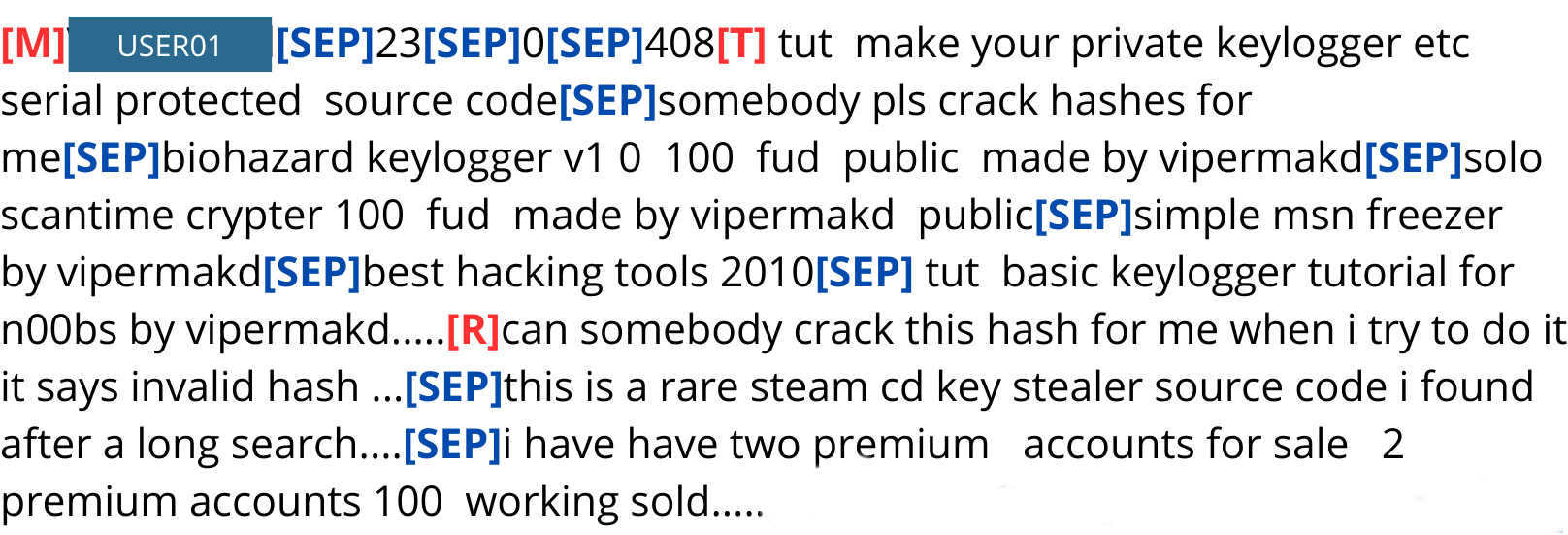}
    \caption{\textbf{Segmented textual user representation.} To preserve structural context for the pretrained language model, user sequences are constructed using specialized demarcating tokens for metadata ([M]), threads ([T]), and replies ([R]), with distinct attributes separated by [SEP] tokens.}
    \label{fig:user_sequence}
\end{figure}

We construct each user profile by concatenating these components into a single textual sequence. This sequence-based approach captures complex user attributes more effectively than traditional feature representations, aligning with findings in recent studies~\cite{amadou2024eurekha,cai2024lmbot}. We encode the metadata as follows:
\begin{center}
\texttt{[M]} \textit{username} \texttt{[SEP]} \textit{reputation} \texttt{[SEP]} \textit{thread\_count} \dots
\end{center}
where \texttt{[M]} marks the start of the metadata segment and \texttt{[SEP]} separates individual attributes. We apply this same schema to threads and replies, introducing each with specific segment tokens. To explicitly distinguish these segments during encoding, we add the special tokens \texttt{[M]} (metadata), \texttt{[T]} (threads), and \texttt{[R]} (replies) to the embedding model's vocabulary. Fig.~\ref{fig:user_sequence} illustrates an example of the resulting user sequence.

\subsection{User Embedding}
\label{section:embedding}
The embedding process transforms variable-length textual content into fixed-length vectors, with dimensionality typically ranging from 256 to 1,536 depending on the model. Approaches range from term-based models, such as Term Frequency-Inverse Document Frequency (TF-IDF)~\cite{106765.106782}, which measures word importance through frequency statistics but lacks semantic depth~\cite{ramos2003using}, to advanced contextual embeddings generated by LLMs. Early word embeddings, such as Word2Vec~\cite{mikolov2013efficient} and GloVe~\cite{pennington2014glove}, captured semantic relationships through co-occurrence patterns but assigned static vectors to words independent of their context. Contextual models such as BERT (Bidirectional Encoder Representations from Transformers) resolved this limitation by producing dynamic embeddings that reflect the surrounding text, thereby significantly improving semantic accuracy~\cite{devlin2019bert}. Subsequent LLMs, such as OpenAI’s GPT, generate general-purpose embeddings~\cite{brown2020language} that capture complex linguistic patterns across a diverse range of applications.

The HADES framework embeds each user sequence into a dense vector using a transformer-based encoder suited for clustering. These embeddings capture both contextual semantics and structural information to represent comprehensive user activity profiles.

Let $S_u = \{M_u, T_u, R_u\}$ denote the textual sequence for user $u$, where $M_u$ is metadata, $T_u$ is thread content, and $R_u$ is reply content. The sequence is tokenized into a sequence of tokens $\{t_1, t_2, \dots, t_n\}$, which includes any required special tokens. The embedding model $f_{\theta}$, parameterized by $\theta$, maps the tokenized sequence to a dense vector:
\[
\mathbf{z}_u = f_{\theta}(S_u) = \text{Pool}\big(\text{Transformer}_{\theta}(\{t_1, t_2, \dots, t_n\})\big) \in \mathbb{R}^d
\]
The $\text{Transformer}_{\theta}$ applies self-attention~\cite{vaswani2017attention} and feed-forward layers to generate contextualized token representations. These representations are then aggregated via a mean pooling operation ($\text{Pool}$)~\cite{reimers2019sentence} into a fixed-length vector. This ensures the resulting embedding $\mathbf{z}_u$ captures both the semantic content and the structural relationships within the user's sequence.

To generate these embeddings, we selected three transformer models that span distinct architectural paradigms and representational trade-offs, enabling a systematic comparison of embedding strategies for underground forum user clustering. Specifically, we evaluate: (1) BERT~\cite{devlin2019bert}, which provides robust bidirectional representations of user sequences; (2) T5~\cite{raffel2020exploring}, a sequence-to-sequence model capable of broad semantic understanding; and (3) All-MiniLM-L6~\cite{sentence-transformers2024}, a distilled model optimized for efficient text representation, making it particularly suited for clustering due to its optimal balance of accuracy and computational speed.

\subsection{User Clustering}
\label{section:clustering}
The user embedding vectors $\mathbf{z}_u \in \mathbb{R}^d$, derived from structured user sequences using transformer-based encoders, serve as the input for clustering. These vectors encapsulate behavioral patterns, facilitating the discovery of community structures comprising semantically similar users within underground forums. Given the high-dimensional, noisy, and potentially non-linearly separable nature of the embedding space, we examined a diverse set of clustering algorithms. Traditional algorithms like \textit{k}-means~\cite{Jin2010} are computationally efficient but require predefined cluster counts and assume spherical distributions. Hierarchical Agglomerative Clustering (AHC)~\cite{mullner2011modern} identifies diverse cluster structures without preset parameters but comes with substantial computational costs. Spectral clustering~\cite{ng2001spectral} uses eigenvalue decomposition to identify non-convex clusters, and Fuzzy C-means~\cite{bezdek1984fcm} supports soft clustering through partial membership. Finally, Hierarchical Density-Based Spatial Clustering of Applications with Noise (HDBSCAN)~\cite{mcinnes2017hdbscan} extends DBSCAN~\cite{ester1996density} by constructing a cluster hierarchy and extracting stable clusters without a predefined count.

Discovering communities within underground forums requires an approach tailored to the unique properties of these embedding spaces. The transformer-derived vectors ($\mathbf{z}_u$) capture high-dimensional behavioral patterns that necessitate an algorithm with three core capabilities: \textbf{(1)}~discovering clusters without a predefined count, reflecting the unknown number of underlying communities; \textbf{(2)}~identifying communities of arbitrary shape and density to accommodate diverse social structures (e.g., small, dense expert groups versus large, diffuse novice communities); and \textbf{(3)}~handling noise to isolate anomalous actors. While the previously mentioned algorithms offer partial solutions, none simultaneously meet all three criteria. For instance, \textit{k}-means~\cite{Jin2010} imposes rigid spherical assumptions and requires a pre-specified cluster count; spectral clustering~\cite{ng2001spectral} is computationally prohibitive for large user bases; and agglomerative methods~\cite{mullner2011modern} lack robustness to the noise and outliers common in this domain.

Therefore, we selected HDBSCAN~\cite{mcinnes2017hdbscan}, which directly aligns with these requirements. HDBSCAN operates without a predefined number of clusters, enabling data-driven exploration of community structures. Its density-based formulation identifies non-spherical clusters of varying densities. Critically, it explicitly models the data distribution to isolate noise, treating anomalous users as outliers rather than forcing them into ill-fitting clusters. HDBSCAN provides a robust method for extracting reliable communities from the complex and noisy ecosystems of underground forums.

\subsection{Cluster Label Generation}
\label{section:label}
HADES generates interpretable labels for each cluster using an open-source LLM. Recent work has demonstrated the effectiveness of LLMs for labeling cybercrime and security-related content at scale. Valeros et al.~\cite{10628591} applied fine-tuned LLMs to translate and annotate underground forum data across languages, while Mischinger et al.~\cite{mischinger2025lost} explored LLM-based classification of non-English cybercrime forums. Beyond forums, Agarwal et al.~\cite{agarwal2025fishing} leveraged language models to classify SMS phishing infrastructure, and Gomez et al.~\cite{GOMEZ2026108313} used similar techniques to clean and label cryptocurrency abuse reports. Building on these methodological foundations, our labeling pipeline follows a structured sampling, generation, and validation protocol designed for the noisy, adversarial language characteristic of underground forums.

\begin{figure}[h]
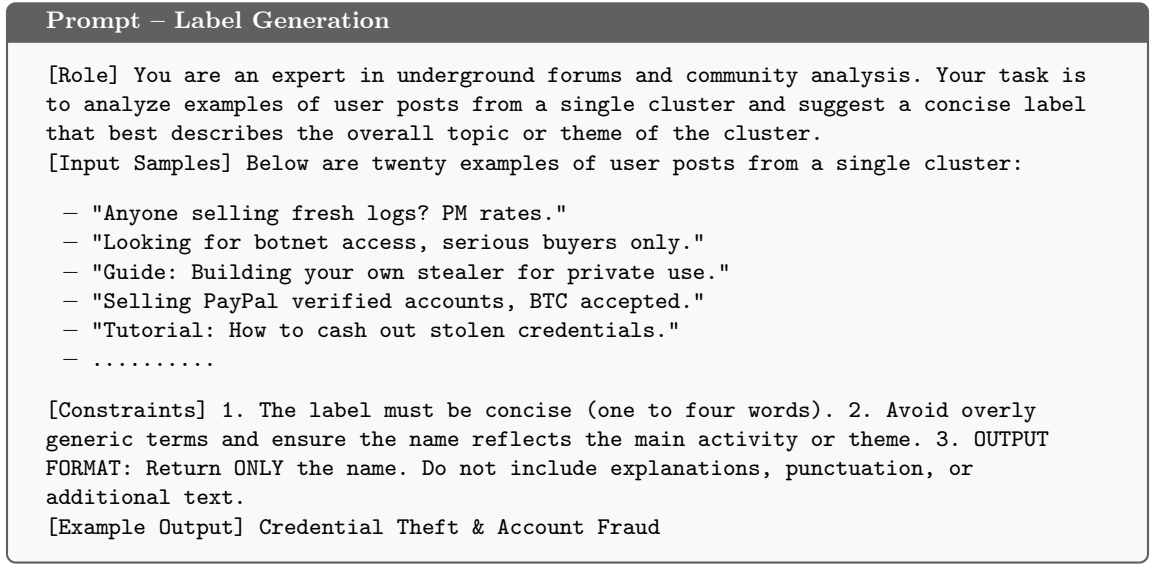

\centering
\begin{tcolorbox}[
    colback=gray!5!white,
    colframe=gray!75!black,
    title=\textbf{Prompt -- Label Generation},
    fontupper=\small\ttfamily,
    boxrule=0.8pt,
]
\textbf{[Role]}
You are an expert in underground forums and community analysis. Your task is to analyze examples of user posts from a single cluster and suggest a concise label that best describes the overall topic or theme of the cluster.

\textbf{[Input Samples]}
Below are twenty examples of user posts from a single cluster:
\begin{itemize}
    \item "Anyone selling fresh logs? PM rates."
    \item "Looking for botnet access, serious buyers only."
    \item "Guide: Building your own stealer for private use."
    \item "Selling PayPal verified accounts, BTC accepted."
    \item "Tutorial: How to cash out stolen credentials."
    \item ..........
\end{itemize}

\textbf{[Constraints]}
1. The label must be concise (one to four words).
2. Avoid overly generic terms and ensure the name reflects the main activity or theme.
3. OUTPUT FORMAT: Return ONLY the name. Do not include explanations, punctuation, or additional text.

\textbf{[Example Output]}
Credential Theft \& Account Fraud
\end{tcolorbox}
\caption{\textbf{Structured prompt template for automated cluster labeling.} Representative samples of user posts from an identified cluster are provided to the LLM.}
\label{fig:prompt_label_gen}
\end{figure}

For each cluster, we randomly sample the textual content of 20 users and prompt the LLM to identify an appropriate thematic category. Text exceeding the model's context window is truncated to the maximum supported length. To assess label consistency, we repeated this sampling process five times per cluster, evaluating 100 distinct users per cluster in total. The prompt used for this task is provided in Fig.~\ref{fig:prompt_label_gen}. To verify the quality and reliability of the generated labels, we conducted a manual validation. One evaluator assessed whether the generated label accurately captured the dominant theme of the sampled content.

\subsection{Evaluation Metrics}
\label{section:evaluation}
To evaluate the communities discovered, we employ four intrinsic metrics that assess two distinct aspects of cluster quality: structural integrity and semantic coherence. Structural integrity measures the geometric separation and compactness of clusters in the embedding space using the Silhouette Score~\cite{silhouette_score}, Calinski–Harabasz Index~\cite{wang2019improved}, and Dunn Index~\cite{bezdek1995cluster}. Semantic coherence~\cite{mimno-etal-2011-optimizing} evaluates whether the communities represent topically consistent user groups. Together, the structural metrics confirm the geometric separation of user behaviors, while semantic coherence verifies that these clusters reflect shared interests—a defining trait of communities in underground forums. We formulate each metric as follows:

\noindent Silhouette Score (SS): This metric quantifies how well a user embedding fits its assigned cluster compared to neighboring clusters, balancing intra-cluster cohesion and inter-cluster separation~\cite{rousseeuw1987silhouettes}. For a user embedding \(\mathbf{e}_u\) in cluster \(C_i\), the score is:
\[
    s(u) = \frac{b(u) - a(u)}{\max\{a(u), b(u)\}}
\]
where \(a(u) = \frac{1}{|C_i| - 1} \sum_{\mathbf{e}_v \in C_i, v \neq u} \|\mathbf{e}_u - \mathbf{e}_v\|_2\) is the average distance to other points in the same cluster, and \(b(u) = \min_{j \neq i} \frac{1}{|C_j|} \sum_{\mathbf{e}_v \in C_j} \|\mathbf{e}_u - \mathbf{e}_v\|_2\) is the average distance to the nearest neighboring cluster. The overall score is the mean \(s(u)\) across all users, ranging from -1 (poor clustering) to 1 (well-separated clusters).

\noindent Semantic Coherence (SC): This metric measures topical alignment by computing the cosine similarity between cluster centroids and topic vectors. For a cluster \(C_i\) with centroid \(\boldsymbol{\mu}_i\) and a set of \(M\) topic embeddings \(\{\mathbf{t}_j\}_{j=1}^M\), the coherence score is: 
\[
    SC_i = \max_{j} \frac{\boldsymbol{\mu}_i \cdot \mathbf{t}_j}{\|\boldsymbol{\mu}_i\|_2 \|\mathbf{t}_j\|_2}
\]
The overall semantic coherence is the average \(SC_i\) across all clusters. Values closer to 1 indicate strong topical alignment, confirming that clusters represent semantically meaningful content.

\noindent Calinski-Harabasz Index (CH): Also known as the Variance Ratio Criterion, this metric evaluates cluster quality by comparing between-cluster dispersion to within-cluster dispersion~\cite{calinski1974dendrite}. It is defined as:
\[
    CH = \frac{\text{tr}(B_k) / (k - 1)}{\text{tr}(W_k) / (N - k)}
\]
where \(B_k=\sum_{i=1}^k |C_i| (\boldsymbol{\mu}_i - \boldsymbol{\mu})(\boldsymbol{\mu}_i - \boldsymbol{\mu})^T\) is the between-cluster covariance matrix, \(W_k = \sum_{i=1}^k \sum_{ \mathbf{e}_u \in C_i} \\(\mathbf{e}_u - \boldsymbol{\mu}_i)(\mathbf{e}_u - \boldsymbol{\mu}_i)^T\) is the within-cluster covariance matrix, \(N\) is the total number of embeddings, \(k\) is the number of clusters, \(\boldsymbol{\mu}\) is the global centroid, and \(\text{tr}(\cdot)\) denotes the trace. Higher values indicate dense, well-defined clusters.

\noindent Dunn Index (DI): This metric compares the minimum inter-cluster distance to the maximum intra-cluster diameter~\cite{dunn1974well}. It is defined as:
\[
    DI = \frac{\min_{i \neq j} d(C_i, C_j)}{\max_{l} \Delta(C_l)}
\]
where \(d(C_i, C_j) = \min_{\mathbf{e}_u \in C_i, \mathbf{e}_v \in C_j} \|\mathbf{e}_u - \mathbf{e}_v\|_2\) is the minimum distance between any two points in clusters \(C_i\) and \(C_j\), and \(\Delta(C_l) = \max_{\mathbf{e}_u, \mathbf{e}_v \in C_l} \|\mathbf{e}_u - \mathbf{e}_v\|_2\) is the diameter of cluster \(C_l\). Higher values indicate well-separated, compact communities. This index is particularly robust for evaluating algorithms like HDBSCAN, which must distinguish dense core communities from noisy outliers.

The Silhouette and Dunn indices assess the geometric compactness and local separation of the partitions. The Calinski-Harabasz Index validates the global structure across varying community sizes, while Semantic Coherence confirms that these structural boundaries align with relevant forum themes.

\section{Results and Discussions}
\subsection{Experimental Setup}
Experiments were conducted on a high-performance computing cluster running RedHat Linux, equipped with an NVIDIA A100 GPU (80 GB VRAM). All implementations were developed in Python 3.8.5. The evaluated pretrained language models embeddings include BERT\footnote{\url{https://huggingface.co/google-bert/bert-base-uncased}}, T5\footnote{\url{https://huggingface.co/google/flan-t5-base}}, and All-MiniLM-L6\footnote{\url{https://huggingface.co/sentence-transformers/all-MiniLM-L6-v2}}, all of which were accessed via the Hugging Face \texttt{transformers} library~\cite{huggingface2024}.

We generated labels using Qwen2.5~\cite{qwen2.5}, a 7.62-billion-parameter decoder-only LLM. This model supports input sequences of up to 128,000 tokens and outputs of up to 8,000 tokens, accommodating long, information-rich contexts. Furthermore, Qwen2.5 supports over 29 languages, which is highly advantageous for analyzing user-generated content in underground forums where discussions frequently span multiple languages, dialects, and writing styles.

To mitigate high-dimensionality effects, all embeddings were reduced to 50 dimensions using UMAP~\cite{mcinnes2018umap} before clustering. We then applied HDBSCAN for clustering, setting \texttt{min\_cluster\_size} to 5, \texttt{min\_samples} to 1, and using the Euclidean distance metric. Clustering quality was evaluated using the Silhouette Score and Calinski-Harabasz Index (via scikit-learn~\cite{pedregosa2011scikit}), as well as the Dunn Index. Each algorithm was executed 10 times for each embedding model, and we report the average results. Our source code is publicly available at \url{https://github.com/osteoner/HADES}.

\subsection{General Clustering Overview}
Table~\ref{tab:best_clustering_results} presents the clustering results for various combinations of forums, embeddings, and the HDBSCAN algorithm. These results reveal a clear hierarchy in embedding performance, demonstrating that model choice significantly affects both clustering quality and the nature of the detected communities.  
BERT embeddings outperform the alternatives across most metrics. For example, BERT achieves the highest Silhouette Score of 0.423 on Cracked and HackForums, indicating well-separated and cohesive clusters. Additionally, BERT achieves the highest Calinski-Harabasz Index (CHI), which measures the ratio of between-cluster to within-cluster dispersion, peaking at 2,502.45 on BreachForums. This strong performance reflects BERT’s pre-training on large, diverse text corpora, which enables it to capture nuanced semantics and subtle distinctions in user language and behavior. Consequently, BERT produces context-aware vector representations that facilitate the identification of coherent user groups. Although All-MiniLM-L6 achieves a notable Silhouette Score of 0.476 on BreachForums, its performance on other forums, similar to that of T5, is moderate, producing clusters that are less geometrically distinct than those of BERT. Nevertheless, both All-MiniLM-L6 and T5 produce clusters with strong semantic coherence and thematically consistent topics. This indicates that while BERT excels at forming mathematically distinct groups, All-MiniLM-L6 and T5 remain effective for thematic grouping despite less separation in the vector space. However, All-MiniLM-L6 exhibits dataset-dependent variability in semantic coherence, a sensitivity potentially linked to its training characteristics.

\begin{figure*}
    \begin{subfigure}[b]{0.325\textwidth}
        \includegraphics[width=\linewidth]{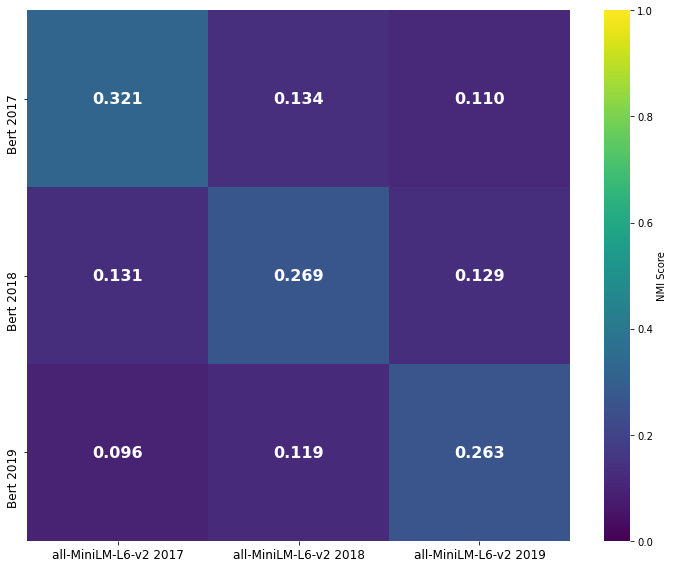}
        \caption{BERT vs All-MiniLM-L6}
        \label{fig:nmi-2017}
    \end{subfigure}
    \hfill
    \begin{subfigure}[b]{0.325\textwidth}
        \includegraphics[width=\linewidth]{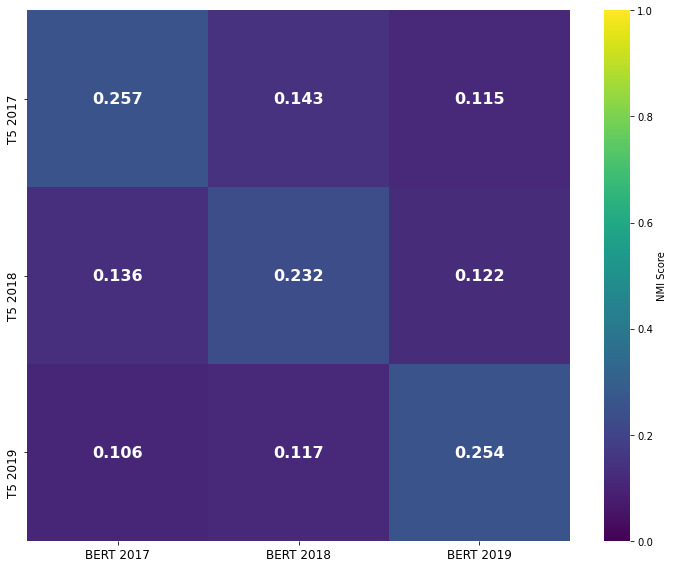}
        \caption{BERT vs T5}
        \label{fig:nmi-2018}
    \end{subfigure}
    \hfill
    \begin{subfigure}[b]{0.325\textwidth}
        \includegraphics[width=\linewidth]{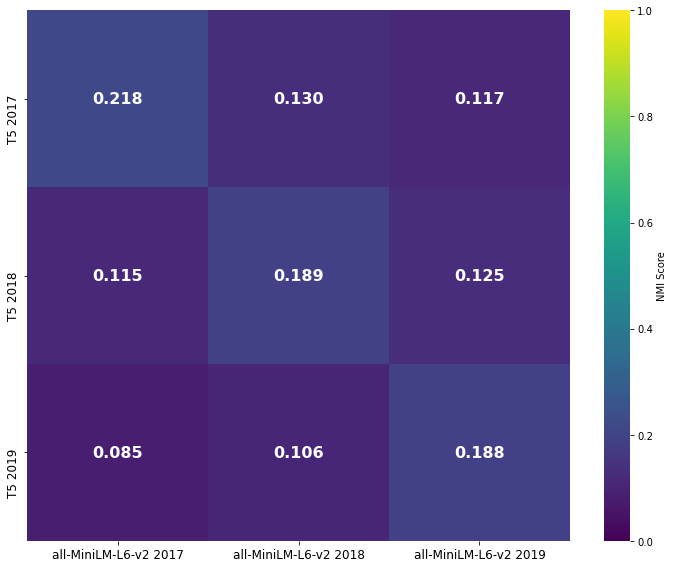}
        \caption{T5 vs All-MiniLM-L6}
        \label{fig:nmi-2019}
    \end{subfigure}

    \caption{\textbf{Structural similarity of community partitions across embedding models.} NMI heatmaps compare the communities generated using BERT, All-MiniLM-L6, and T5 embeddings on HackForums (2017–2019).}
    \label{fig:community-comparaison}
\end{figure*}

We quantitatively compare the community partitions generated by different pretrained language model embeddings using the Normalized Mutual Information (NMI) score. NMI is a standard information-theoretic metric that measures the similarity between two sets of communities on a scale from 0 to 1, where a high score indicates similar community structures. Fig.~\ref{fig:community-comparaison} shows pairwise NMI comparisons of communities detected using BERT, All-MiniLM-L6, and T5 over three years (2017–2019) on HackForums. We observe consistently high NMI scores between communities detected with BERT and All-MiniLM-L6 during this period, particularly in 2017, which aligns with their similar architectures. In contrast, T5 achieves lower NMI scores when compared to BERT and All-MiniLM-L6, indicating that T5 captures different semantic features and produces distinct user groupings.  

The choice of embedding model significantly shapes the structure of the detected communities. For instance, applying BERT embeddings to HackForums identified 56 distinct communities, compared to 39 with T5. This difference highlights the varying representational capacities of the embeddings; BERT’s ability to capture fine-grained distinctions facilitates the detection of more specialized subgroups. Consequently, the choice of embedding model is a crucial methodological decision that shapes the scale, characteristics, and interpretation of the identified communities, as well as their underlying social structures.

\begin{infobox}
\textbf{Lessons learned.} Our analysis demonstrates that the architecture and training paradigms of pretrained language model embeddings significantly influence community detection. Models sharing the same architecture and training data, such as BERT and All-MiniLM-L6, tend to produce similar community structures, though BERT consistently performs best. Across forums, BERT achieves the highest Silhouette Score (0.423) and Calinski-Harabasz Index (2,502.45), achieving well-separated, semantically coherent clusters that capture subtle user differences. While All-MiniLM-L6 and T5 also form meaningful communities, their clusters are less distinct, and their performance is less consistent. Notably, BERT identifies more fine-grained communities, such as 56 in HackForums, compared to 42 with T5, highlighting its capacity to capture nuanced user subgroups. Furthermore, the similarity between communities detected by BERT and All-MiniLM-L6 declines over time as the forum evolves, reflecting the impact of a changing user base and content diversity.
\end{infobox}

\begin{table*}
\centering
\caption{\textbf{Quantitative evaluation of clustering performance across different pretrained embedding models.}}
\label{tab:best_clustering_results}
\resizebox{1\textwidth}{!}{%
\begin{tblr}{
  cell{1}{3} = {c},
  cell{1}{4} = {c},
  cell{1}{5} = {c},
  cell{1}{6} = {c},
  cell{1}{7} = {c},
  cell{2}{1} = {r=3}{},
  cell{2}{3} = {c},
  cell{2}{4} = {c},
  cell{2}{5} = {c},
  cell{2}{6} = {c},
  cell{2}{7} = {c},
  cell{3}{3} = {c},
  cell{3}{4} = {c},
  cell{3}{5} = {c},
  cell{3}{6} = {c},
  cell{3}{7} = {c},
  cell{4}{3} = {c},
  cell{4}{4} = {c},
  cell{4}{5} = {c},
  cell{4}{6} = {c},
  cell{4}{7} = {c},
  cell{5}{1} = {r=3}{},
  cell{5}{7} = {c},
  cell{6}{7} = {c},
  cell{7}{7} = {c},
  cell{8}{1} = {r=3}{},
  cell{8}{7} = {c},
  cell{9}{7} = {c},
  cell{10}{7} = {c},
  hline{1-2,5,8,11} = {-}{},
  hline{5,8,11} = {2-7}{},
}
\textbf{Dataset} & \textbf{Embedding} & \textbf{SS}    & \textbf{CHI}      & \textbf{Dunn}  & \textbf{SC}    & \textbf{ \# Communities} \\
HackForums       & BERT               & \textbf{0.423} & \textbf{351.34}   & 0.209          & \textbf{0.992} & 56 \\
                 & All-MiniLM-L6      & 0.318          & 90.59             & \textbf{0.286} & 0.875          & 42 \\
                 & T5                 & 0.155          & 86.69             & 0.270          & 0.792          & 39 \\
Cracked          & BERT               & \textbf{0.423} & \textbf{336.79}   & 0.196          & \textbf{0.994} & 86 \\
                 & All-MiniLM-L6      & 0.320          & 83.71             & 0.232          & 0.894          & 76 \\
                 & T5                 & 0.304          & 157.31            & \textbf{0.277} & 0.710          & 65 \\
BreachForums     &    BERT    & \textbf{0.476} & 219.53            & 0.074          & \textbf{0.965}          & 52 \\
                 & All-MiniLM-L6               & 0.442          & \textbf{2502.45}  & 0.177          & 0.894 & 48 \\
                 & T5                 & 0.315          & 1222.92           & \textbf{0.235} & 0.882          & 31
\end{tblr}
}
\end{table*}

\subsection{Community Evolution Analysis}
To evaluate community evolution in underground forums, we compared HADES, which utilizes BERT embeddings, with the Louvain algorithm \cite{blondel2008fast} used in prior studies \cite{MANATOVA2024114271}. We conducted this comparison using data from HackForums over three years (2017–2020), tracking 16,111 active users. Our analysis indicates that while both methods detect broad community structures, HADES more effectively identifies hidden communities and assigns descriptive labels.

\begin{figure*}
    \centering

    \begin{subfigure}[b]{0.32\textwidth}
        \includegraphics[width=\linewidth]{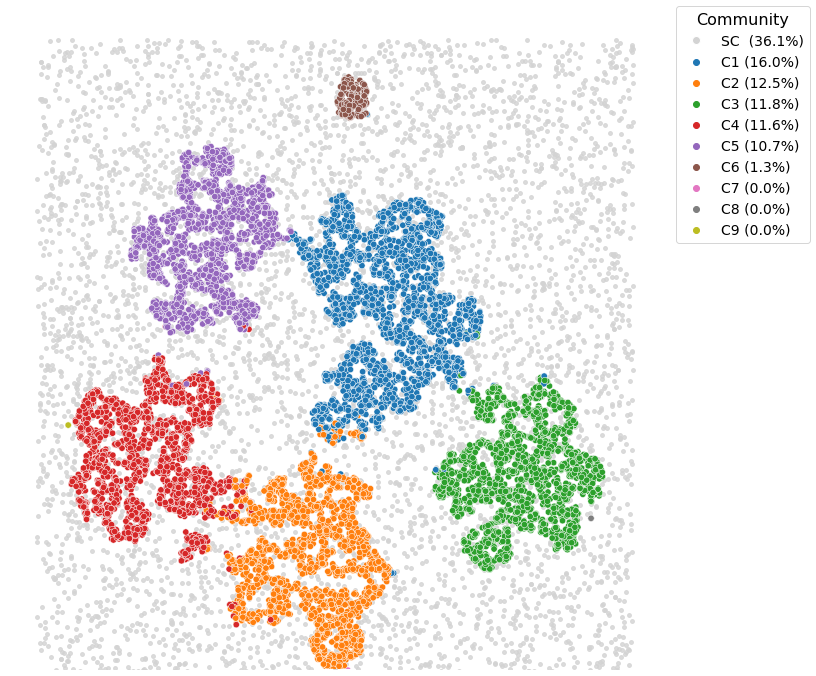}
        \caption{Graph-based 2017}
        \label{fig:louvain-2017}
    \end{subfigure}
    \hfill
    \begin{subfigure}[b]{0.32\textwidth}
        \includegraphics[width=\linewidth]{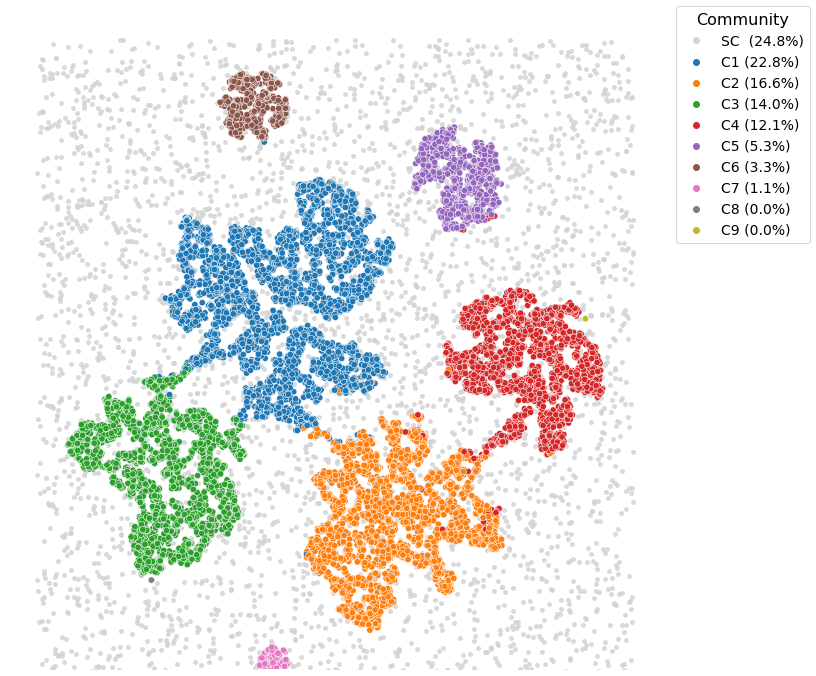}
        \caption{Graph-based 2018}
        \label{fig:louvain-2018}
    \end{subfigure}
    \hfill
    \begin{subfigure}[b]{0.32\textwidth}
        \includegraphics[width=\linewidth]{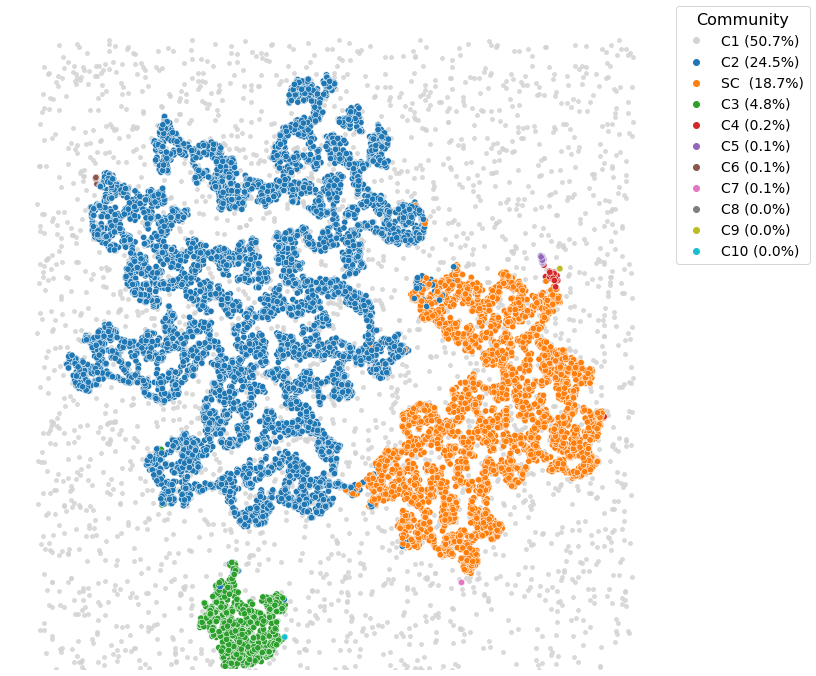}
        \caption{Graph-based 2019}
        \label{fig:louvain-2019}
    \end{subfigure}

    \begin{subfigure}[b]{0.32\textwidth}
        \includegraphics[width=\linewidth]{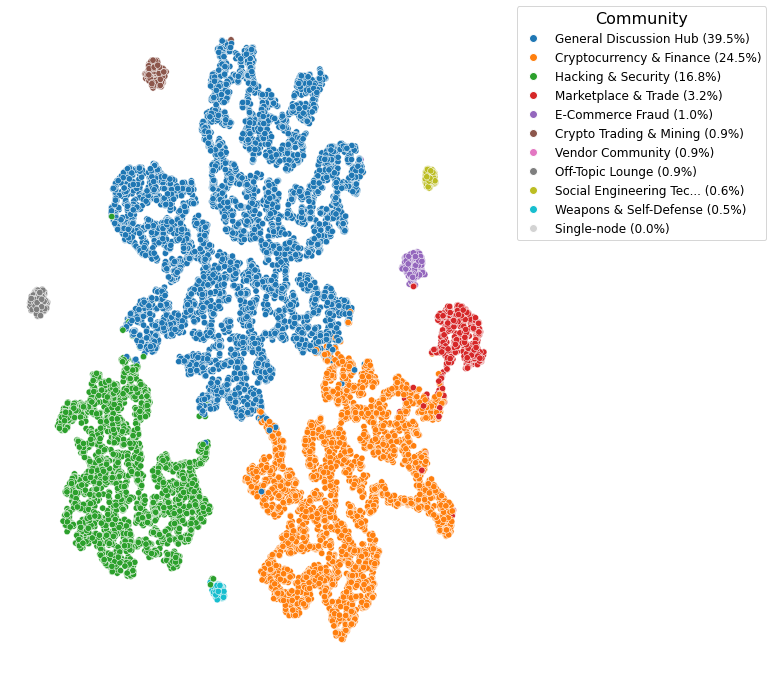}
        \caption{HADES 2017}
        \label{fig:HADES-2017}
    \end{subfigure}
    \hfill
    \begin{subfigure}[b]{0.32\textwidth}
        \includegraphics[width=\linewidth]{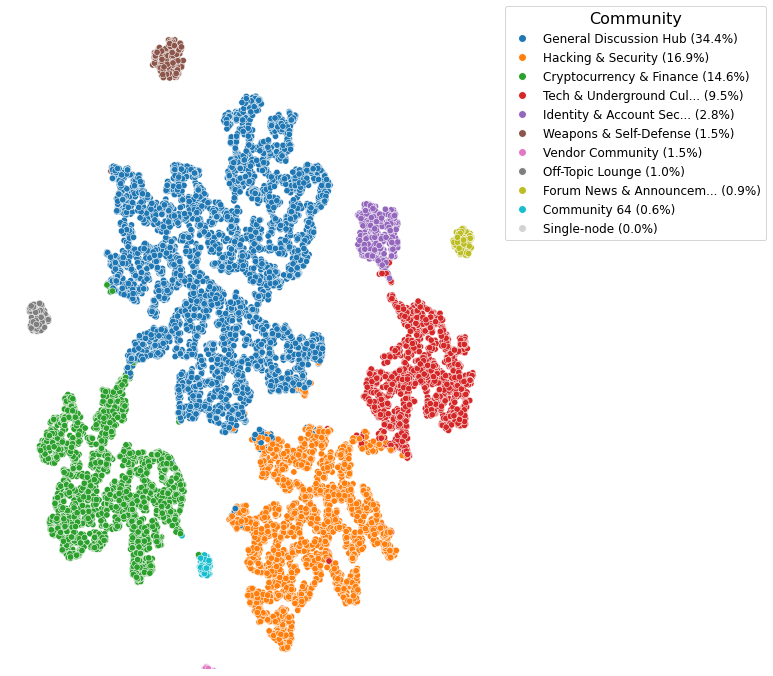}
        \caption{HADES 2018}
        \label{fig:HADES-2018}
    \end{subfigure}
    \hfill
    \begin{subfigure}[b]{0.32\textwidth}
        \includegraphics[width=\linewidth]{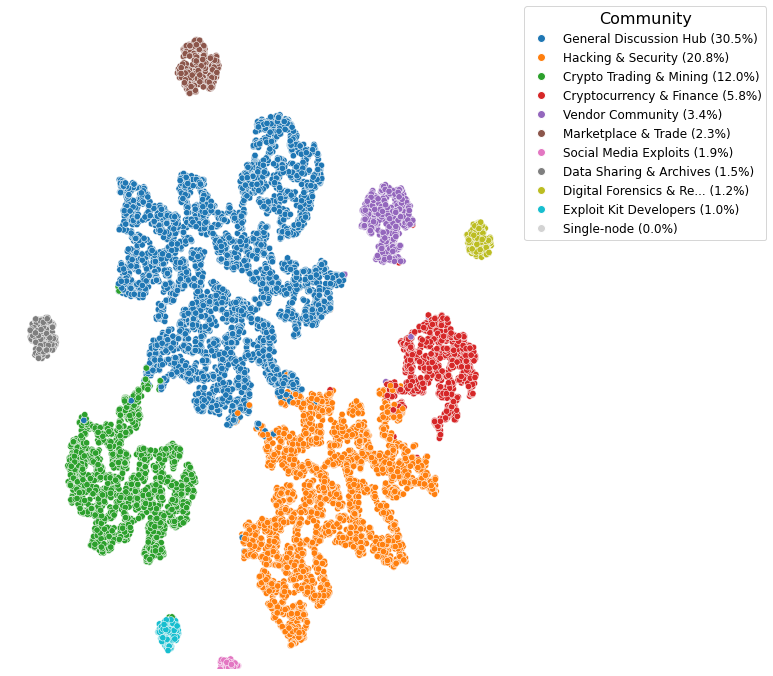}
        \caption{HADES 2019}
        \label{fig:HADES-2019}
    \end{subfigure}

    \caption{\textbf{Longitudinal t-SNE visualization of community evolution on HackForums (2017–2019).} The graph-based approach initially isolates a large number of users into single-node communities (gray points).}
    \label{fig:community-evolution-combined}
\end{figure*}

We constructed annual social graphs by linking users who contributed to threads initiated by others, modeling implicit social connections consistent with previous research~\cite{pete2020social, huang2021hackerrank, MANATOVA2024114271}. Users posting in the same thread were considered socially connected, enabling us to capture interaction dynamics and information flow. For HADES, we represented each user’s annual activity as a sequence and generated corresponding embeddings. Both the Louvain algorithm and HADES were applied separately to the user data for each year, allowing for a direct temporal comparison between the two approaches. 
As shown in Fig.~\ref{fig:louvain-2017},\ref{fig:louvain-2019}, the Louvain algorithm primarily identified large, dominant communities and did not assign descriptive labels. Furthermore, in 2017, the Louvain algorithm classified 36.1\% of users (6,366 out of 16,111) into 6,335 single-user communities. This fragmentation reflects the sparsity of the underlying interaction graph~\cite{Traag2018FromLT}. In contrast, HADES produced no single-user communities (0\%) during the same period, demonstrating its capability to resolve hidden group affiliations and apply informative labels despite sparse network interactions.

By 2019, user interactions had evolved significantly:
\begin{itemize}
\item \textbf{Integration of Isolated Users:} Of the 5,829 users isolated in 2017 (Fig.\ref{fig:louvain-2017}), 4,365 (74.88\%) joined multi-user communities by 2019 (Fig.\ref{fig:louvain-2019}), with over 13\% of previously isolated users integrating into the largest community. HADES had already grouped these users into communities such as the "General Discussion Hub" in 2017 (Fig.\ref{fig:HADES-2017}), demonstrating its capacity to detect latent ties before they manifest structurally.
\item \textbf{Community Transitions:} Among the 10,326 users located in multi-user communities in 2017, 7,391 (71.58\%) transitioned into the primary community by 2019, often migrating between groups with overlapping interests.
\item \textbf{Emergence of Community C2:} A notable community, labeled C2 by the Louvain algorithm in 2019 (Fig.\ref{fig:HADES-2019}), consisted of 3,966 users. Of these, 3,416 (86.13\%) had been isolated in 2017 (Fig.\ref{fig:louvain-2017}). HADES identified the foundation of this community as early as 2017 (Fig.\ref{fig:HADES-2017}), grouping 52.94\% of these users into a "Hacking \& Security" community. This early detection underscores HADES’s consistency in tracking community evolution over time.
\end{itemize}

\begin{figure}
    \centering
    \includegraphics[width=0.6\linewidth]{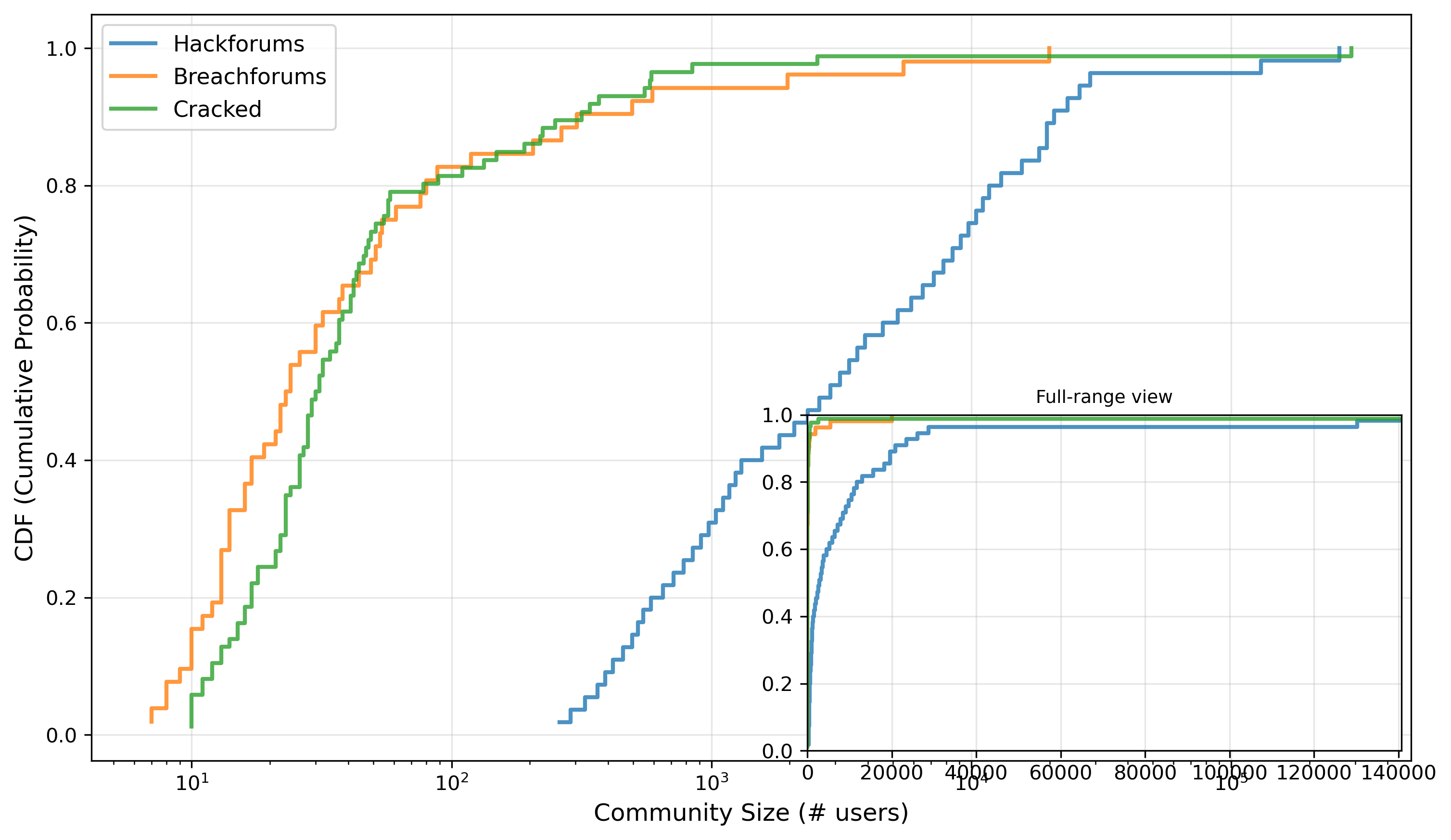}
    \caption{\textbf{Cumulative Distribution Function (CDF) of community sizes across three underground forums.} A few general-interest hubs contain the majority of users, while over 80\% of the communities are specialized micro-communities with fewer than 100 members.}
    \label{fig:cdf-forums}
\end{figure}

\begin{infobox}
\textbf{Lessons learned.} While the Louvain algorithm effectively detects dominant communities, it is limited by sparse interaction graphs, often resulting in numerous single-node communities. In contrast, HADES identifies both prominent and hidden communities, detecting latent group affiliations up to a year before they manifest in graph-based structural analyses. Furthermore, the model generates consistent, descriptive labels that capture the thematic evolution of user discussions over time. This capacity for early detection enables the proactive tracking of community evolution, providing a robust approach for analyzing and anticipating dynamic shifts within online communities.
\end{infobox}

\subsection{Dataset-Specific Insights}
Using HADES with BERT embeddings, we qualitatively analyzed communities across three underground forums. This analysis identified distinct thematic hubs focused on General Discussion, Hacking, Gaming, Coding, Technology, Markets, and Finance. Table~\ref{tab:communities_big_small} lists the 10 largest and 10 smallest communities.
A clear pattern emerged: the largest communities on each forum are broad, general-interest hubs with the highest membership. Examples include the "General Discussion Hub" on HackForums (260,470 members), "Lounge" on Cracked (289,820 members), and "General Discussion" on BreachForums (19,995 members). Other major communities center on high-engagement topics such as trading, hacking, and cryptocurrency. In contrast, the smallest communities are highly specialized and often dedicated to illicit activities. As shown in Fig.~\ref{fig:cdf-forums}, more than 80\% of communities on Cracked and BreachForums have fewer than 100 users. For example, the "Ransomware Groups" community on BreachForums and the "Dark Market" community on Cracked consist of only 7 to 16 members each. These micro-communities facilitate collaboration on niche underground services, such as phishing kit development or unauthorized data trading.
HackForums, by comparison, exhibits a substantially higher level of activity, containing 11 times more posts and 9 times more threads than the other two forums combined (Table~\ref{tab:forum_datasets}). Consequently, even its smallest communities comprise approximately 260 members. Fig.~\ref{fig:cdf-forums} shows that 80\% of HackForums communities fall within the range of 260 to 1,000 users. Such communities are difficult to detect using traditional graph-based methods that rely solely on user interactions, because their formation is typically driven by shared interests, may take years to develop, and often requires a high-profile member to attract others. By leveraging user interests, our framework can not only detect these communities but also predict their emergence up to a year in advance.
Examining the "Exploit Kit Developers" community on HackForums, we found that although members share similar interests, they often possess varying levels of expertise. For example, an advanced user might post a detailed tutorial on building a Python remote access trojan (RAT), providing the full source code and techniques for evading detection, while offering advanced features via private message. Conversely, a novice user (e.g., "CrypterNoob") might seek advice on applying crypters to obfuscate RATs and avoid antivirus detection. Although both users participate in the same community and engage with similar topics, the disparity in their technical proficiency is significant. This contrast demonstrates that community membership reflects shared interests rather than uniform technical skill or experience.

\begin{infobox}
\textbf{Lessons learned.} Our qualitative analysis shows that underground communities exhibit both broad and highly specialized structures. Large, general-interest hubs attract the majority of users, while numerous smaller micro-communities often focus on specific illicit activities. This distinction highlights the limitations of traditional graph-based community detection methods, which rely primarily on user interactions. Such methods often fail to identify emerging communities, particularly those built around shared interests, for years after their formation, or until a high-profile member triggers broader growth. In contrast, HADES can detect and anticipate the emergence of these communities up to a year in advance. Furthermore, we observe that community membership is driven by shared interests rather than uniform expertise, with members ranging from novices seeking guidance to experts sharing advanced techniques.
\end{infobox}

\begin{table*}[ht]
\centering
\caption{\textbf{Size and thematic diversity of the top-10 largest and smallest detected communities.} Results demonstrate that while the largest communities act as broad, general-discussion hubs (e.g., "General Discussion Hub" with 260k users), the smallest groups are highly specialized micro-communities dedicated to niche, high-risk illicit operations (e.g., "Exploit Kit Developers", "Initial Access Brokers").}
\label{tab:communities_big_small}
\begin{adjustbox}{max width=\textwidth} 
\begin{tabular}{@{}clrlrlr@{}}
\toprule
\textbf{Rank} & \textbf{HackForums} & \textbf{Size} 
              & \textbf{BreachForums} & \textbf{Size} 
              & \textbf{Cracked} & \textbf{Size} \\
\midrule
\multicolumn{7}{c}{\textbf{Largest Communities}} \\
\midrule
1  & General Discussion Hub        & 260,470 & General Discussion          & 19,995 & Lounge      & 289,820 \\
2  & Marketplace \& Trade          & 130,235 & Exchange Forum              & 5,495  & Hacktivism         & 2,561 \\
3  & Crypto Trading \& Mining      & 28,652  & Leeching \& Piracy          & 1,967  & Fiesta Exploit      & 844 \\
4  & Cryptocurrency \& Finance     & 26,047  & Dark Web Pharmacies \& Med. & 593    & Thanks Humor        & 587 \\
5  & Fraud \& Scam Reports         & 23,442  & Leaked Data \& CC Dumps     & 495    & Graphic Design      & 580 \\
6  & Data Sharing \& Archives      & 20,838  & Android RAT \& Exploits     & 303    & Hack Discussion     & 554 \\
7  & Vendor Community              & 19,535  & Anime Recommendations       & 206    & Bitcoin Leeches     & 369 \\
8  & Tech \& Underground Culture   & 19,535  & Business Data Marketplace   & 119    & Underground Hacks   & 316 \\
9  & Off-Topic Lounge              & 18,233  & Electronic Music Forum      & 61     & Leak Community      & 340 \\
10 & Hacking \& Security           & 15,628  & Movie Sharing \& Streaming  & 54     & Internet Humor      & 250 \\
\midrule
\multicolumn{7}{c}{\textbf{Smallest Communities}} \\
\midrule
1  & Credential Stuffing Groups    & 547  & Ransomware Groups            & 16 & Dork Quality        & 16 \\
2  & Darknet Escrow Services       & 521  & Child Exploitation \& Porn   & 16 & Bitcoin Forum       & 16 \\
3  & Exploit Kit Developers        & 495  & Hacked IoT Devices           & 10 & Xbox Sharing        & 12 \\
4  & Blackhat SEO Networks         & 456  & Social Dynamics \& Ethics    & 10 & Crypto Passerby     & 11 \\
5  & Spam-as-a-Service Crews       & 417  & Hacking \& Credential Trading& 10 & Underground Deals   & 11 \\
6  & Cashout Services              & 391  & Twitter Data Leak            & 9  & Crypto Forums       & 13 \\
7  & Bulletproof Hosting           & 365  & Markets \& Sales             & 8  & AI Discourse        & 10 \\
8  & Initial Access Brokers        & 326  & Tools \& Hacking Techniques  & 8  & Dark Market         & 10 \\
9  & Malware-as-a-Service          & 287  & Unauthorized Access Systems  & 7  & Bitcoin Forum       & 10 \\
10 & Phishing Kit Developers       & 260  & Underground Markets \& Deals & 7  & Dark Market         & 10 \\
\bottomrule
\end{tabular}
\end{adjustbox} 
\end{table*}

\subsection{Practical Implications}
HADES offers several practical advantages for CTI operations. First, it detects communities engaged in high-risk activities, such as exploit kit development and Malware-as-a-Service—without requiring a pre-constructed interaction graph. Unlike graph-based methods, which depend on long observation windows to establish reliable interaction networks, our framework identifies emerging communities earlier by leveraging semantic similarity alone. This enables analysts to monitor illicit groups, anticipate evolving threats, and develop countermeasures before attacks mature.

Second, the communities identified by HADES can be enriched through integration with complementary frameworks. Event-level detection systems such as EventHunter~\cite{EchChammakhy2025EventHunterDC} could link detected communities to specific threat incidents (e.g., zero-day disclosures, malware releases), providing analysts with both the \textit{who} (community membership) and the \textit{what} (specific events). Similarly, cybersecurity named entity recognition models such as CyberNER~\cite{EchChammakhy2025CyberNER} could extract structured indicators of compromise from community content, enabling a direct path from community detection to actionable intelligence.

\section{Limitations and Future Work}
\label{limitations}
A primary limitation of this study is the computational overhead of large-scale experiments. Specifically, generating dense vector embeddings with models like BERT for millions of users before clustering is resource-intensive. These constraints restricted our evaluation to smaller models, hindering a comprehensive analysis of state-of-the-art LLMs. Future work should explore efficiency improvements to enable the evaluation of larger models. Second, evaluating our approach is constrained by the lack of ground-truth data. Because baseline labels are unavailable, we relied on manual qualitative verification to assess community boundaries and mitigate the risk of LLM hallucinations during label generation. Developing annotated benchmark datasets would facilitate rigorous quantitative evaluation. We also note a conceptual distinction in how communities are defined and detected in this study. Traditional graph methods detect "structural communities" based on explicit collaboration networks, which require observable, direct user interactions. HADES identifies "semantic communities" as latent topic clusters composed of individuals united by shared interests, technical vocabulary, and behavioral patterns. For instance, grouping a novice seeking tutorials with an advanced malware developer indicates strong thematic alignment, but it does not guarantee an existing collaborative relationship. Identifying these semantic cohorts, therefore, serves as a proxy for structural community formation, allowing analysts to anticipate collaborative threat networks before explicit interactions appear in the network graph. 

Third, our evaluation is limited to English-language forums. Underground forums operate in many languages, including Russian, Chinese, Arabic, and Portuguese, and the preprocessing pipeline currently filters out non-Latin characters, which would remove potentially informative content in multilingual contexts. Extending HADES to non-English forums is a natural direction for future work. 

Finally, expanding our limited search for HDBSCAN parameters could refine community boundaries and improve the framework's effectiveness.

\section{Ethical considerations}\label{ethical}
Our research uses the CrimeBB academic dataset~\cite{Pastrana2018CrimeBB}, which consists of historical, publicly available text collected from underground forums. We accessed these data under a formal Data Use Agreement (DUA) with the Cambridge Cybercrime Centre, which prohibits the redistribution of raw data. Given the sensitive nature of this domain, we adhered to strict ethical standards. Because our study is entirely non-interventional and relies exclusively on public posts, it involved no interaction with forum users. 

\section{Conclusion}
This paper introduces HADES, an unsupervised framework designed to detect both dominant and hidden threat communities within cybercriminal underground forums. By using pretrained language models to generate semantic embeddings from user textual representations, our approach captures shared interests and latent behavioral patterns that traditional graph-based methods often overlook. By clustering users based on this semantic similarity and assigning topic labels to the resulting clusters, HADES identifies specific threat communities to support CTI. Our evaluation across HackForums, Cracked, and BreachForums demonstrates that BERT-based embeddings consistently achieve high cluster coherence. This allows the framework to successfully isolate small, highly specialized micro-communities, often comprising fewer than 100 users, that typically evade conventional detection. Furthermore, because users engage with specific topics before establishing explicit structural connections, tracking these semantic patterns enables HADES to identify emerging communities up to a year earlier than graph-based methods. This framework provides a scalable, proactive tool for CTI, allowing security analysts to monitor and anticipate evolving threats.

\section*{Acknowledgements}
The authors gratefully acknowledge the Cambridge Cybercrime Centre, UK, for providing access to the CrimeBB dataset, which made this research possible.

\bibliographystyle{plain}
\bibliography{references}

\begin{thebibliography}{10}

\bibitem{agarwal2025fishing}
Sharad Agarwal, Antonis Papasavva, Guillermo Suarez-Tangil, and Marie Vasek.
\newblock Fishing for smishing: Understanding sms phishing infrastructure and strategies by mining public user reports.
\newblock In {\em Proceedings of the 2025 ACM Internet Measurement Conference}, pages 223--241, 2025.

\bibitem{amadou2024eurekha}
Abdoul Nasser~Hassane Amadou, Anas Motii, Saida Elouardi, and El~Houcine Bergou.
\newblock Eurekha: Enhancing user representation for key hackers identification in underground forums.
\newblock In {\em 2024 IEEE 23rd International Conference on Trust, Security and Privacy in Computing and Communications (TrustCom)}, pages 387--398, 2024.

\bibitem{amadou2024hc}
Abdoul Nasser~Hassane Amadou, Anas Motii, and Mohammed Jouhari.
\newblock Hc-hackerrank: Identifying key hackers in cybercrime social network forums.
\newblock In {\em 2024 7th International Conference on Advanced Communication Technologies and Networking (CommNet)}, pages 1--8, 2024.

\bibitem{bezdek1984fcm}
James~C Bezdek, Robert Ehrlich, and William Full.
\newblock Fcm: The fuzzy c-means clustering algorithm.
\newblock {\em Computers \& Geosciences}, 10(2-3):191--203, 1984.

\bibitem{bezdek1995cluster}
James~C Bezdek and Nikhil~R Pal.
\newblock Cluster validation with generalized dunn's indices.
\newblock In {\em Proceedings of the 2nd New Zealand International Two-Stream Conference on Artificial Neural Networks and Expert Systems}, pages 190--193, 1995.

\bibitem{blondel2008fast}
Vincent~D Blondel, Jean-Loup Guillaume, Renaud Lambiotte, and Etienne Lefebvre.
\newblock Fast unfolding of communities in large networks.
\newblock {\em Journal of Statistical Mechanics: Theory and Experiment}, (10):P10008, 2008.

\bibitem{brown2020language}
Tom Brown, Benjamin Mann, Nick Ryder, Melanie Subbiah, Jared~D Kaplan, Prafulla Dhariwal, Arvind Neelakantan, Pranav Shyam, Girish Sastry, Amanda Askell, et~al.
\newblock Language models are few-shot learners.
\newblock {\em Advances in Neural Information Processing Systems}, 33:1877--1901, 2020.

\bibitem{cabrero2021methodology}
Jos{\'e} Cabrero-Holgueras and Sergio Pastrana.
\newblock A methodology for large-scale identification of related accounts in underground forums.
\newblock {\em Computers \& Security}, 111:102489, 2021.

\bibitem{cai2024lmbot}
Zijian Cai, Zhaoxuan Tan, Zhenyu Lei, Zifeng Zhu, Hongrui Wang, Qinghua Zheng, and Minnan Luo.
\newblock {LMBot: Distilling Graph Knowledge into Language Model for Graph-Less Deployment in Twitter Bot Detection}.
\newblock In {\em Proceedings of the 17th {ACM} International Conference on Web Search and Data Mining}, pages 57--66, 2024.

\bibitem{calinski1974dendrite}
T.~Caliński and J.~Harabasz.
\newblock A dendrite method for cluster analysis.
\newblock {\em Communications in Statistics - Theory and Methods}, 3(1):1--27, 1974.

\bibitem{devlin2019bert}
Jacob Devlin, Ming-Wei Chang, Kenton Lee, and Kristina Toutanova.
\newblock {Bert: Pre-training of deep bidirectional transformers for language understanding}.
\newblock In {\em Proceedings of the 2019 conference of the North American chapter of the association for computational linguistics: Human Language Technologies, vol. 1}, pages 4171--4186, 2019.

\bibitem{dunn1974well}
J.~C. Dunn.
\newblock Well-separated clusters and optimal fuzzy partitions.
\newblock {\em Journal of Cybernetics}, 4(1):95--104, 1974.

\bibitem{EchChammakhy2025CyberNER}
Yasir Ech-Chammakhy, Anas Motii, Anass Rabii, Oussama Azrara, and Jaafar Chbili.
\newblock { CyberNER: A Harmonized STIX Corpus for Cybersecurity Named Entity Recognition }.
\newblock In {\em 2025 IEEE 24th International Conference on Trust, Security and Privacy in Computing and Communications (TrustCom)}, pages 2190--2197, 2025.

\bibitem{EchChammakhy2025EventHunterDC}
Yasir Ech-Chammakhy, Anas Motii, Anass Rabii, and Jaafar Chbili.
\newblock Eventhunter: Dynamic clustering and ranking of security events from hacker forum discussions.
\newblock {\em Proceedings of the 28th International Symposium on Research in Attacks, Intrusions and Defenses}, pages 552--565, 2025.

\bibitem{ester1996density}
Martin Ester, Hans-Peter Kriegel, J{\"o}rg Sander, and Xiaowei Xu.
\newblock Density-based spatial clustering of applications with noise.
\newblock In {\em Proceedings of the 2nd International Conference on Knowledge Discovery and Data Mining}, volume 240, 1996.

\bibitem{GOMEZ2026108313}
Gibran Gomez, Kevin {van Liebergen}, Davide Sanvito, Giuseppe Siracusano, Roberto Gonzalez, and Juan Caballero.
\newblock Clean up the mess: Addressing data pollution in cryptocurrency abuse reporting services.
\newblock {\em Future Generation Computer Systems}, 179:108313, 2026.

\bibitem{huang2021hackerrank}
Cheng Huang, Yongyan Guo, Wenbo Guo, and Ying Li.
\newblock {HackerRank: Identifying key hackers in underground forums}.
\newblock {\em International Journal of Distributed Sensor Networks}, 17(5), 2021.

\bibitem{huang2019topic}
Shin-Ying Huang and Tao Ban.
\newblock A topic-based unsupervised learning approach for online underground market exploration.
\newblock In {\em Proceedings of 18th IEEE International Conference on Trust, Security and Privacy in Computing and communications/13th IEEE International Conference on Big Data Science and Engineering (TrustCom/BigDataSE)}, pages 208--215, 2019.

\bibitem{huang2016exploring}
Shin-Ying Huang and Hsinchun Chen.
\newblock Exploring the online underground marketplaces through topic-based social network and clustering.
\newblock In {\em 2016 IEEE Conference on Intelligence and Security Informatics (ISI)}, pages 145--150, 2016.

\bibitem{huggingface2024}
{Hugging Face}.
\newblock {Hugging Face – The AI community building the future}, 2024.
\newblock Accessed: 2025-05-27.

\bibitem{Jin2010}
Xin Jin and Jiawei Han.
\newblock {\em K-Means Clustering}, pages 563--564.
\newblock 2010.

\bibitem{li2024comprehensive}
Jiakang Li, Songning Lai, Zhihao Shuai, Yuan Tan, Yifan Jia, Mianyang Yu, Zichen Song, Xiaokang Peng, Ziyang Xu, Yongxin Ni, et~al.
\newblock A comprehensive review of community detection in graphs.
\newblock {\em Neurocomputing}, 600:128169, 2024.

\bibitem{maharana2022review}
Kiran Maharana, Surajit Mondal, and Bhushankumar Nemade.
\newblock A review: Data pre-processing and data augmentation techniques.
\newblock {\em Global Transitions Proceedings}, 3(1):91--99, 2022.

\bibitem{MANATOVA2024114271}
Dalyapraz Manatova, Charles DeVries, and Sagar Samtani.
\newblock Understand your shady neighborhood: An approach for detecting and investigating hacker communities.
\newblock {\em Decision Support Systems}, 184:114271, 2024.

\bibitem{mcinnes2017hdbscan}
Leland McInnes, John Healy, Steve Astels, et~al.
\newblock {HDBSCAN}: Hierarchical density based clustering.
\newblock {\em Journal of Open Source Software}, 2(11):205, 2017.

\bibitem{mcinnes2018umap}
Leland McInnes, John Healy, and James Melville.
\newblock {UMAP}: Uniform manifold approximation and projection for dimension reduction.
\newblock {\em arXiv preprint arXiv:1802.03426}, 2018.

\bibitem{mikolov2013efficient}
Tomas Mikolov, Kai Chen, Greg Corrado, and Jeffrey Dean.
\newblock Efficient estimation of word representations in vector space.
\newblock {\em arXiv preprint arXiv:1301.3781}, 2013.

\bibitem{mimno-etal-2011-optimizing}
David Mimno, Hanna Wallach, Edmund Talley, Miriam Leenders, and Andrew McCallum.
\newblock Optimizing semantic coherence in topic models.
\newblock In {\em Proceedings of the 2011 Conference on Empirical Methods in Natural Language Processing}, pages 262--272, 2011.

\bibitem{mischinger2025lost}
Mariella Mischinger, Jack Hughes, Fedor Vitiugin, Sergio Pastrana, Alice Hutchings, and Guillermo Suarez-Tangil.
\newblock Lost in translation: Analyzing non-english cybercrime forums.
\newblock In {\em 2025 APWG Symposium on Electronic Crime Research}, pages 1--16, 2025.

\bibitem{mullner2011modern}
Daniel M{\"u}llner.
\newblock Modern hierarchical, agglomerative clustering algorithms.
\newblock {\em arXiv preprint arXiv:1109.2378}, 2011.

\bibitem{newman2004finding}
Mark~EJ Newman and Michelle Girvan.
\newblock Finding and evaluating community structure in networks.
\newblock {\em Physical Review E}, 69(2):026113, 2004.

\bibitem{ng2001spectral}
Andrew Ng, Michael Jordan, and Yair Weiss.
\newblock On spectral clustering: Analysis and an algorithm.
\newblock {\em Advances in Neural Information Processing Systems}, 14, 2001.

\bibitem{paracha2023sus}
Anum~Atique Paracha, Junaid Arshad, and Muhammad~Mubashir Khan.
\newblock Sus you’re sus!—identifying influencer hackers on dark web social networks.
\newblock {\em Computers \& Electrical Engineering}, 107:108627, 2023.

\bibitem{Pastrana2018CrimeBB}
Sergio Pastrana, Daniel~R Thomas, Alice Hutchings, and Richard Clayton.
\newblock {CrimeBB}: Enabling cybercrime research on underground forums at scale.
\newblock In {\em Proceedings of the 2018 World Wide Web Conference}, pages 1845--1854, 2018.

\bibitem{pedregosa2011scikit}
Fabian Pedregosa, Gael Varoquaux, Alexandre Gramfort, Vincent Michel, Bertrand Thirion, Olivier Grisel, Mathieu Blondel, Peter Prettenhofer, Ron Weiss, Vincent Dubourg, et~al.
\newblock {Scikit-learn: Machine Learning in Python. Journal of Machine Learning Research, 12}.
\newblock 2011.

\bibitem{pennington2014glove}
Jeffrey Pennington, Richard Socher, and Christopher~D Manning.
\newblock Glove: Global vectors for word representation.
\newblock In {\em Proceedings of the 2014 Conference on Empirical Methods in Natural Language Processing}, pages 1532--1543, 2014.

\bibitem{pete2022postcog}
Ildiko Pete, Jack Hughes, Andrew Caines, Anh~V Vu, Harshad Gupta, Alice Hutchings, Ross Anderson, and Paula Buttery.
\newblock Postcog: A tool for interdisciplinary research into underground forums at scale.
\newblock In {\em 2022 IEEE European Symposium on Security and Privacy Workshops}, pages 93--104, 2022.

\bibitem{pete2020social}
Ildiko Pete, Jack Hughes, Yi~Ting Chua, and Maria Bada.
\newblock A social network analysis and comparison of six dark web forums.
\newblock In {\em 2020 IEEE European Symposium on Security and Privacy Workshops}, pages 484--493, 2020.

\bibitem{petukhova2022textcl}
Alina Petukhova and Nuno Fachada.
\newblock {TextCL: A Python package for NLP preprocessing tasks}.
\newblock {\em SoftwareX}, 19:101122, 2022.

\bibitem{pourhabibi2021darknetexplorer}
Tahereh Pourhabibi, Kok-Leong Ong, Booi~H Kam, and Yee~Ling Boo.
\newblock {DarkNetExplorer (DNE): Exploring dark multi-layer networks beyond the resolution limit}.
\newblock {\em Decision Support Systems}, 146:113537, 2021.

\bibitem{raffel2020exploring}
Colin Raffel, Noam Shazeer, Adam Roberts, Katherine Lee, Sharan Narang, Michael Matena, Yanqi Zhou, Wei Li, and Peter~J Liu.
\newblock Exploring the limits of transfer learning with a unified text-to-text transformer.
\newblock {\em Journal of Machine Learning Research}, 21(140):1--67, 2020.

\bibitem{ramos2003using}
Juan Ramos et~al.
\newblock Using tf-idf to determine word relevance in document queries.
\newblock In {\em Proceedings of the First Instructional Conference on Machine Learning}, volume 242, pages 29--48, 2003.

\bibitem{reimers2019sentence}
Nils Reimers and Iryna Gurevych.
\newblock {Sentence-BERT: Sentence Embeddings using Siamese BERT-Networks}.
\newblock In {\em Proceedings of the 2019 Conference on Empirical Methods in Natural Language Processing}, pages 3982--3992, 2019.

\bibitem{Rios2012KDD}
Sebasti\'{a}n~A. R\'{\i}os and Ricardo Mu\~{n}oz.
\newblock {Dark Web portal overlapping community detection based on topic models}.
\newblock In {\em Proceedings of the ACM SIGKDD Workshop on Intelligence and Security Informatics}, 2012.

\bibitem{rousseeuw1987silhouettes}
Peter~J. Rousseeuw.
\newblock Silhouettes: A graphical aid to the interpretation and validation of cluster analysis.
\newblock {\em Journal of Computational and Applied Mathematics}, 20:53--65, 1987.

\bibitem{samtani2015exploring}
Sagar Samtani, Ryan Chinn, and Hsinchun Chen.
\newblock Exploring hacker assets in underground forums.
\newblock In {\em IEEE International Conference on Intelligence and Security Informatics}, pages 31--36, 2015.

\bibitem{sentence-transformers2024}
{Sentence-Transformers}.
\newblock {all-MiniLM-L6-v2}.
\newblock \url{https://huggingface.co/sentence-transformers/all-MiniLM-L6-v2}, 2024.

\bibitem{sentinelone2026}
{SentinelOne}.
\newblock {Key Cybersecurity Statistics for 2026}.
\newblock \url{https://www.sentinelone.com/cybersecurity-101/cybersecurity/cyber-security-statistics/}, 2026.

\bibitem{silhouette_score}
Ketan~Rajshekhar Shahapure and Charles Nicholas.
\newblock Cluster quality analysis using silhouette score.
\newblock In {\em 2020 IEEE 7th International Conference on Data Science and Advanced Analytics}, pages 747--748, 2020.

\bibitem{868688}
Jianbo Shi and J.~Malik.
\newblock Normalized cuts and image segmentation.
\newblock {\em IEEE Transactions on Pattern Analysis and Machine Intelligence}, 22(8):888--905, 2000.

\bibitem{106765.106782}
Karen Sparck~Jones.
\newblock {\em A statistical interpretation of term specificity and its application in retrieval}, page 132–142.
\newblock 1988.

\bibitem{qwen2.5}
Qwen Team.
\newblock Qwen2.5: A party of foundation models.
\newblock \url{https://qwenlm.github.io/blog/qwen2.5/}, 2024.

\bibitem{Traag2018FromLT}
Vincent~Antonio Traag, Ludo Waltman, and Nees~Jan van Eck.
\newblock From louvain to leiden: guaranteeing well-connected communities.
\newblock {\em Scientific Reports}, 9, 2019.

\bibitem{10628591}
Veronica Valeros, Anna Sirokova, Carlos Catania, and Sebastian Garcia.
\newblock { Towards Better Understanding of Cybercrime: The Role of Fine-Tuned LLMs in Translation }.
\newblock In {\em 2024 IEEE European Symposium on Security and Privacy Workshops}, pages 91--99, 2024.

\bibitem{vaswani2017attention}
Ashish Vaswani, Noam Shazeer, Niki Parmar, Jakob Uszkoreit, Llion Jones, Aidan~N Gomez, {\L}ukasz Kaiser, and Illia Polosukhin.
\newblock Attention is all you need.
\newblock In {\em Advances in Neural Information Processing Systems}, volume~30, pages 5998--6008, 2017.

\bibitem{wang2019improved}
Xu~Wang and Yusheng Xu.
\newblock {An improved index for clustering validation based on Silhouette index and Calinski-Harabasz index}.
\newblock In {\em IOP Conference Series: Materials Science and Engineering}, volume 569, page 052024, 2019.

\bibitem{wilson2024identifying}
Lydia Wilson, Viet~Anh Vu, Ildik{\'o} Pete, and Yi~Ting Chua.
\newblock {Identifying and Collecting Public Domain Data for Tracking Cybercrime and Online Extremism}.
\newblock In {\em Open Source Investigations in the Age of Google}, pages 281--301. 2024.

\bibitem{XU2022848}
Yijia Xu, Yong Fang, Cheng Huang, and Zhonglin Liu.
\newblock {HGHAN: Hacker group identification based on heterogeneous graph attention network}.
\newblock {\em Information Sciences}, 612:848--863, 2022.

\bibitem{XU2024111587}
Yijia Xu, Yong Fang, Cheng Huang, Zhonglin Liu, and Weipeng Cao.
\newblock Hacker group identification based on dynamic heterogeneous graph node update.
\newblock {\em Applied Soft Computing}, 158:111587, 2024.

\end{thebibliography}
\end{document}